\documentclass[conference,final]{IEEEtran}
\usepackage{epsfig,endnotes}
\usepackage{tugraz_defaults}
\usepackage{csquotes}
\let\labelindent\relax
\usepackage{enumitem}
\usepackage{flushend}
\usepackage{subcaption}
\usepackage{lstautogobble}

\usepackage{tcolorbox}
\tcbset{colback=gray!10!white}

\renewcommand{\paragraph}[1]{\vspace{0.0cm}\noindent\textbf{#1}\ }

\begin{document}
\title{Domain Page-Table Isolation}
\date{}

\author{\IEEEauthorblockN{Claudio Canella\IEEEauthorrefmark{1},
Andreas Kogler\IEEEauthorrefmark{1},
Lukas Giner\IEEEauthorrefmark{1},
Daniel Gruss\IEEEauthorrefmark{1} and
Michael Schwarz\IEEEauthorrefmark{2}}
\IEEEauthorblockA{\IEEEauthorrefmark{1}Graz University of Technology\\ Email: firstname.lastname@iaik.tugraz.at}
\IEEEauthorblockA{\IEEEauthorrefmark{2}CISPA Helmholtz Center for Information Security\\
Email: michael.schwarz@cispa.saarland}}

\newcommand{\syscall}{syscall\xspace}
\newcommand{\Syscall}{Syscall\xspace}
\newcommand{\syscalls}{syscalls\xspace}
\newcommand{\Syscalls}{Syscalls\xspace}

\newcommand{\Approach}{DPTI\xspace}
\newcommand{\ApproachHide}{DPTI-Stash\xspace}
\newcommand{\ApproachFreeze}{DPTI-Freeze\xspace}

\maketitle

\begin{abstract}
Modern applications often consist of different security domains that require isolation from each other. 
While several solutions exist, most of them rely on specialized hardware, hardware extensions, or require less-efficient software instrumentation of the application.

In this paper, we propose Domain Page-Table Isolation (\Approach), a novel mechanism for hardware-enforced security domains that can be readily used on commodity off-the-shelf CPUs.
\Approach uses two novel techniques for dynamic, time-limited changes to the memory isolation at security-critical points, called memory \textit{freezing} and \textit{stashing}.
We demonstrate the versatility and efficacy of \Approach in two scenarios:
First, \Approach \textit{freezes} or \textit{stashes} memory to support faster and more fine-grained \syscall filtering than state-of-the-art seccomp-bpf.
With the provided memory-safety guarantees, \Approach can even securely support deep argument filtering, such as string comparisons.
Second, \Approach \textit{freezes} or \textit{stashes} memory to efficiently confine potentially untrusted SGX enclaves, outperforming existing solutions by \SI{14.6}{\percent}-\SI{22}{\percent} while providing the same security guarantees.
Our results show that \Approach is a viable mechanism to isolate domains within applications using only existing mechanisms available on modern CPUs, without relying on special hardware instructions or extensions. 
\end{abstract}

\section{Introduction} \label{sec:intro}

Memory isolation is a vital primitive to ensure the security of modern systems. 
However, this isolation is not only necessary between different applications, but it becomes more and more important within applications as well. 
Often, applications consist of multiple security domains that should be isolated from each other. 
Such in-process isolation ensures that vulnerabilities in one domain cannot easily affect a different domain. 
The separation of user and kernel space is a well-established isolation mechanism, trusted execution environments (TEEs) such as Intel SGX or ARM TrustZone are more recent examples.
The interface and memory model of TEEs share similarities with the kernel-user-space boundary, where the enclave is considered trusted, but the rest of the application is not.
Other new ISA extensions, such as Intel MPK~\cite{Intel_vol3}, allow setting up multiple security domains within applications to efficiently control access permissions for different memory ranges.
In all these cases, the isolation is configured in software and enforced by the hardware. 

However, while the isolation must provide strong security guarantees, it is still necessary that the domains can communicate with each other. 
This interface between low-privilege and higher-privileged domains is often exploited.
For instance, the \syscall interface between unprivileged applications and the kernel is often used for privilege escalation attacks~\cite{Szekeres2013sok}.
Hence, state-of-the-art isolation techniques such as seccomp-bpf~\cite{Edge2015seccomp} provide developers with the ability to block \syscalls, reducing the kernel's attack surface in case of exploited applications. 
While this works well for blocking \syscalls, the performance overhead can increase linearly with the length of the blocklist~\cite{Hromatka2018,Tizen2014}.
Still, while securing this interface as well as possible is important, there is currently no option to create sophisticated filters.
For example, it is not possible to block the \texttt{exec} \syscall for all but a hard-coded set of applications.
This lack of support for deep argument filtering is not merely a matter of implementation but stems from the current design of the filtering mechanism~\cite{Edge2019,Edge2020}.

Existing isolation solutions are often limited to very specific scenarios, require specialized hardware, or source- or binary-level instrumentation of the application.
For example, the hardware isolation of TEEs can only be applied to enclaves.
However, in practice, it is infeasible to run all software in enclaves.
Enclaves have limitations, such as the inability to directly interact with the kernel, and are only available on a subset of CPUs.
Their asymmetric trust model was exploited in recent work~\cite{Schwarz2019SGXMalware,Weiser2019SGXJail}, requiring mitigations with high overheads or special hardware features.
Proposals for domain isolation require hardware extensions or modifications as well~\cite{Schrammel2020Donky,Weiser2019SGXJail,Frassetto2018}. 
Moreover, hardware extensions such as Intel MPK are not broadly available, \eg MPK is only available on Xeon Scalable CPUs and some 10th-generation Intel CPUs. 
Furthermore, an attacker can disable Intel MPK security domains after gaining code execution within the protected application, as preventing this is not in scope for Intel MPK~\cite{Vahldiek2018erim}. 
Hence, while these proposed isolation mechanisms are effective, they \emph{cannot be used on commodity CPUs}.

In this paper, we introduce the notion of dynamic, time-limited changes to the memory isolation at security-critical points, a security-domain mechanism we call \textbf{\Approach} (Domain Page-Table Isolation).
As \Approach-based solutions rely only on the existing hardware-enforced memory protection of the memory-management unit (MMU), we can implement memory isolation for sandboxing and domain-isolation scenarios on commodity, off-the-shelf CPUs.
Still, \Approach has the same page granularity as other approaches that require dedicated hardware or hardware extensions~\cite{Vahldiek2018erim,Schrammel2020Donky,Weiser2019SGXJail,Park2019libmpk}. 
Similar to KPTI~\cite{LWN_kpti,Gruss2017KASLR}, \Approach uses efficient page-table modifications to create security domains that are temporarily inaccessible from another domain.
However, in contrast to KPTI, \Approach does not necessarily have to maintain a second set of page tables for every process, and it has to solve additional challenges such as multiple mappings to one physical page.
In the remainder of this work, we will use \textit{\Approach} as shorthand for both the concept and its proof-of-concept implementations.

We evaluate \Approach on Intel CPUs ranging from 2015 (for SGX support) to 2018.
However, as \Approach only relies on the MMU, it runs on all CPUs with virtual-memory support, in contrast to previous works~\cite{Frassetto2018,Mogosanu2018,Vahldiek2018erim,Schrammel2020Donky,Koning2017,Hedayati2019hodor,Park2019libmpk}.
Contrary to software fault isolation (SFI), we do not use any instrumentation of the application.
With \Approach, it is even possible to add stricter \syscall filters to existing (legacy) applications via a wrapper application. 

We demonstrate the versatility of \Approach in two scenarios, enhanced \syscall filtering and SGX protection domains, leading to high-performance, feature-rich solutions. 
In these, \Approach relies on two new techniques: memory \textit{freezing} and \textit{stashing}.

In our first application of \Approach, we present \syscall filtering that is both faster and more fine-grained than seccomp-bpf.
\Approach temporarily \textit{freezes} or \textit{stashes} the memory ranges of complex \syscall parameters from the untrusted user-space application on a \syscall, preventing modification while the kernel performs the \syscall.
While this intuitively sounds straightforward, it requires solving multiple challenges, including process creation and replacement, multithreading support, alias mappings, and alternative memory-access interfaces. 
However, by solving these challenges, \Approach-based filtering can evaluate complex parameters, such as strings, without introducing time-of-check-to-time-of-use (TOCTOU) vulnerabilities into the \syscall interface~\cite{Schwarz2018DF,Edge2020,Edge2019}.
Additionally, simple, non-argument-inspecting \Approach-based allow and reject filters only incur an overhead of \SI{22}{\percent} compared to seccomp-bpf's \SI{33.9}{\percent}.
Especially filter-heavy applications benefit from our implementation since the overhead does not depend on the number of filters, as is the case for seccomp~\cite{Hromatka2018,Tizen2014}.

In the second use case, \Approach improves performance and security of SGX enclave confinement~\cite{Schwarz2019SGXMalware,Weiser2019SGXJail}. 
\Approach prevents the rewriting of host memory from SGX, again by \textit{stashing} or \textit{freezing} host memory, making it inaccessible to enclaves.
However, this is not trivial: switching to and from an enclave requires some host pages.
We address this challenge by securing SGX entries and exits with a special code bridge page.
Thus, we mitigate host impersonation for arbitrary \syscall execution by a malicious or hijacked SGX enclave~\cite{Schwarz2019SGXMalware}. 
While \Approach outperforms existing SGX enclave confinement solutions~\cite{Weiser2019SGXJail} by \SI{14.6}{\percent}-\SI{22}{\percent}, the most important advantage is that \Approach works on commodity systems without hardware changes and maintains full compatibility with existing enclaves. 

\Approach is an efficient and easily implementable solution for isolating security domains in various scenarios.
Due to its software-only implementation without specific ISA-extension dependencies~\cite{Frassetto2018,Mogosanu2018,Vahldiek2018erim,Schrammel2020Donky,Koning2017,Hedayati2019hodor,Park2019libmpk}, it can be readily used on commodity, off-the-shelf hardware.

To summarize, we make the following contributions:
\begin{compactenum}
  \item We introduce \Approach,\footnote{Prototype can be found at \url{https://github.com/domain-isolation/DPTI}.} an MMU-based isolation mechanism inspired by page-table isolation, enabling fine-grained security domains and security policies. 
  \item We use \Approach to implement extended \syscall filtering, enabling sophisticated argument-inspecting filter rules not supported by seccomp or AppArmor. 
  \item We show that \Approach can replace existing SGX enclave confinement solutions, isolating unmodified untrusted enclaves on commodity hardware.
  \item We thoroughly evaluate the security and performance of \Approach and show that it provides higher performance and security than previous solutions in both case studies.
\end{compactenum}

\textbf{Outline.}
\cref{sec:background} discusses background.
\cref{sec:idea} describes the high-level idea and threat model.
\cref{sec:case-studies} details the two case studies for \Approach.
\cref{sec:evaluation} evaluates security and performance of \Approach.
\cref{sec:discussion} discusses future and related work.
We conclude in \cref{sec:conclusion}.

\section{Background}\label{sec:background}
This section introduces topics required for this work.

\subsection{Sandboxing}
Sandboxing provides an extra layer of security by strictly controlling the resources an application can access~\cite{Prevelakis2001sandboxing,Goldberg1996secure}. 
In many cases, sandboxing is a last line of defense that assumes that the sandboxed application was already exploited.
Thus, a sandbox should drastically limit the impact of an exploit. 
Existing sandboxes typically restrict access to the network and file system and limit the available \syscalls.
Sandboxes are widespread in browsers~\cite{Firefox2019sandbox,Reis2019SiteIsolation,Firefox2019fission} and on mobile operating systems~\cite{Android2017seccomp,AndroidAppSandbox}.
Linux provides sandboxing via the SELinux~\cite{SELinuxFAQ} and AppArmor~\cite{AppArmor} frameworks. 

\subsection{Linux Seccomp}
A vital part of sandboxes is the ability to restrict the \syscall interface. 
The \syscall interface provides functionality from the operating system to user-space applications. 
An exploit with unrestricted access to the \syscall interface can read, write, and execute files on the system.
In the worst case, the \syscall interface itself is exploitable, leading to a privilege escalation~\cite{Kemerlis2014,Kemerlis2015,Kemerlis2012}.
Hence, sandboxes try to minimize the number of exposed \syscalls to a minimum required for the application to work.
Secure Computing (seccomp)~\cite{Edge2015seccomp} is a \syscall filter that is integrated into the Linux kernel. 
An application can specify allowed \syscalls, and the kernel will block all others. 
In addition to the \syscall itself, seccomp can define filters for integer parameters. 
Filters based on the content of strings or structures are not supported, as seccomp cannot dereference parameters~\cite{Edge2019,Edge2020}.

Due to the complexity of limiting the \syscall interface, several approaches to automatically generate \syscall filters have been published recently~\cite{Ghavamnia2020,Canella2021chestnut,DeMarinis2020}.
The common approach is to dynamically or statically analyze an application to detect all required \syscalls. 

\subsection{Runtime Attacks}
Many security vulnerabilities are caused by memory safety violations.
Typical memory safety violations, such as buffer overflows, enable attackers to modify an application in an unintended way~\cite{Szekeres2013sok}.
In many cases, attackers try to overwrite code pointers to hijack the control flow. 
On modern systems, data is typically not executable, and thus an attacker cannot inject so-called shellcode into an exploited application~\cite{Szekeres2013sok}.
As a result, exploits fall back to reusing existing program parts and diverting the control flow to so-called \textit{gadgets}~\cite{Nergal2001ret2libc}. 
Shacham~\cite{Shacham2007} generalized such control-flow-hijacking attacks using gadgets as return-oriented programming (ROP). 
By chaining multiple gadgets, it is generally possible to build arbitrary exploits. 
In addition to such control-flow-hijacking that overwrite pointers~\cite{Shacham2007,Checkoway2010JOP,Lan2015loop,Goktas2014COP,Schuster2015COOP}, data-only attacks~\cite{Rogowski2017,Ispoglou2018} can also violate memory safety.

\paragraph{Race conditions} are a type of vulnerability where a data structure is accessed in parallel, and the actual order of accesses affects the correctness of the program. 
A special type of race condition are time-of-check-to-time-of-use (TOCTOU) vulnerabilities.
A TOCTOU bug exists if a memory location is accessed multiple times, and an attacker can manipulate the data between the access.
If such a TOCTOU bug is exploitable, it is also called a double-fetch bug~\cite{Wang2017DF,Schwarz2018DF}. 
Double-fetch bugs are especially dangerous in the \syscall interface as they are hard to detect~\cite{Wang2017DF} and often relatively easy to exploit~\cite{Schwarz2018DF}. 

\subsection{Memory Isolation}
Memory isolation has been used to provide security for a long time.
Segmentation and paging are two of the most well-known approaches to isolate memory on x86.
The operating system configures the memory, and the CPU then enforces this configuration.
While segmentation is no longer used to enforce access permissions on x86-64, security researchers continued to propose information hiding techniques based on this legacy feature~\cite{Backes2014oxymoron,Kuznetsov2014CPI,Lu2015aslrguard}.
Modern systems rely on paging to translate the virtual addresses of a process to physical addresses through the help of page tables.
Page tables also contain access permissions which determine whether a page is read-only or user-space accessible.

Other memory isolation techniques use newer hardware features such as Extended Page Tables (EPT)~\cite{Koning2017,Liu2015Isolation}, Software Guard Extensions (SGX)~\cite{Frassetto2017}, and Memory Protection Keys (MPK)~\cite{Vahldiek2018erim,Schrammel2020Donky,Koning2017,Hedayati2019hodor,Park2019libmpk}.
EPTs facilitate memory virtualization, while SGX allows protecting code and data even from a compromised operating system.
MPK introduces a new register containing a protection key and allows a developer to associate memory with such a key.
The key itself is stored in the page tables, and an access is only allowed if the current register key matches the one stored in the page table.

\subsection{SGX}
Intel SGX~\cite{Intel_vol3} is a trusted execution environment (TEE) introduced with the Intel Skylake microarchitecture in 2015.
Using this instruction-set extension, applications can be divided into an untrusted and a trusted part. 
The trusted part, the so-called enclave, is integrity- and confidentiality-protected by the CPU. 
Thus, even in the case of a malicious operating system, the content of the enclave is protected. 
Enclaves are hosted by ordinary, untrusted applications.
Both enclave and untrusted application run in the same virtual address space. 
While the hardware prevents outside access to the address range of the enclave, the enclave can access all memory of the host application. 
The SGX threat model assumes that the entire software stack is malicious. 
However, there is no consideration that an enclave might be malicious. 
This asymmetry enables enclave malware that can impersonate the host application to execute arbitrary \syscalls~\cite{Schwarz2019SGXMalware,Weiser2019SGXJail}.
Enclaves themselves cannot execute \syscalls. 
Instead, enclaves communicate with the host application via ECALLs and OCALLs. 
After loading the enclave, the host application can call secure enclave functions using ECALLs via a call gate, similar to \syscalls. 
If the enclave wants to use functionality from the operating system, such as a \syscall, it has to use an OCALL to call into the host application. 
The ECALL/OCALL interface is defined by the developer at compile time. 

\section{High-Level Idea \& Threat Model}\label{sec:idea}

\begin{figure}[t]
  \begin{subfigure}[b]{0.45\hsize}
  \resizebox{\hsize}{!}{
    \usetikzlibrary{positioning}
\usetikzlibrary{shapes.geometric, arrows}

\tikzstyle{mapping} = [rectangle, minimum width=1.82cm, minimum height=1.5cm,text centered, draw=black,fill=black!2]
\tikzstyle{shared} = [rectangle, minimum width=1.82cm, minimum height=0.5cm,text centered, draw=black,fill=black!2]

\tikzstyle{arrow} = [thick,->,>=stealth,shorten >=4pt,shorten <=4pt]

\begin{tikzpicture}
  \begin{scope}
    \begin{scope}
    \node[draw,mapping] at (0, 0) {Domain A};
    \node[draw,shared] (shared_trusted_left) at (0, -1) {Shared};
    \node[draw,mapping] at (0, -2) {Domain B};
    \end{scope}

    \begin{scope}[shift={(3,0)}]
    \node[draw,mapping] at (0, 0) {Domain A};
    \node[draw,shared] (shared_trusted_right) at (0, -1) {Shared};
    \node[draw,mapping] at (0, -2) {Domain B};
    \end{scope}

    \draw[arrow] (shared_trusted_left.east) -- (shared_trusted_right.west) node[midway,above] {\parbox{1cm}{\centering switch}} node[midway,below] {\centering domain};
  \end{scope}
\end{tikzpicture}
    }
    \caption{Domain A view}
    \label{fig:trusted_view}
  \end{subfigure}
  ~
  \begin{subfigure}[b]{0.45\hsize}
  \resizebox{\hsize}{!}{
    \usetikzlibrary{positioning}
\usetikzlibrary{shapes.geometric, arrows}

\tikzstyle{mapping} = [rectangle, minimum width=1.82cm, minimum height=1.5cm,text centered, draw=black,fill=black!2]
\tikzstyle{shared} = [rectangle, minimum width=1.82cm, minimum height=0.5cm,text centered, draw=black,fill=black!2]
\tikzstyle{arrow} = [thick,->,>=stealth,shorten >=4pt,shorten <=4pt]

\begin{tikzpicture}
  \begin{scope}
    \begin{scope}
    \node[draw,mapping,pattern=north east lines,pattern color=red!60] at (0, 0) {Domain A};
    \node[draw,shared] (shared_untrusted_left) at (0, -1) {Shared};
    \node[draw,mapping] at (0, -2) {Domain B};
    \end{scope}

    \begin{scope}[shift={(3,0)}]
    \node[draw,mapping,pattern=north east lines,pattern color=red!60] at (0, 0) {Domain A};
    \node[draw,shared,pattern=north east lines,pattern color=red!60] (shared_untrusted_right) at (0, -1) {Shared};
    \node[draw,mapping] at (0, -2) {Domain B};
    \end{scope}

    \draw[arrow] (shared_untrusted_left.east) -- (shared_untrusted_right.west) node[midway,above] {\parbox{1cm}{\centering switch}} node[midway,below] {\centering domain};
  \end{scope}
\end{tikzpicture}
    }
    \caption{Domain B view}
    \label{fig:untrusted_view}
  \end{subfigure}
  \caption{The shared address space and the different views of it.
  No change for Domain A upon switching while parts are stashed or frozen (red pattern) for Domain B with \Approach.}
  \label{fig:overview}
\end{figure}
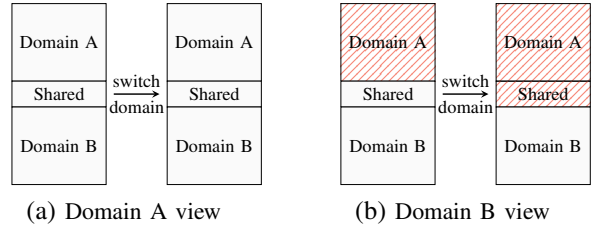

On a high level, \Approach provides software-defined memory protection for more fine-grained sandboxing policies, such as deep argument filtering.
This protection of memory regions can present itself in two different variants: read-only protection or entirely inaccessible.
We refer to the former as \ApproachFreeze and to the latter as \ApproachHide.
Either variant guarantees that the memory cannot be modified by an untrusted domain.

\Cref{fig:overview} shows the idea of \Approach.
Two security domains share one virtual address space and use a dedicated memory region for communication, \ie for passing data across the security boundary. 
In many cases, while there is a defined memory region for the communication, Domain A has access to the entire Domain B. 
As both domains can execute code in parallel, Domain A cannot ensure the atomicity of the accessed data in this memory region.
Hence, Domain B could modify the data while Domain A accesses it, potentially leading to data corruption or TOCTOU vulnerabilities.
\Approach solves this problem by protecting the content of this memory region as long as Domain A accesses it. 
Any modification from Domain B is simply delayed until Domain A finishes its execution. 

Such a scenario is common for operating systems, where the kernel (Domain A) is typically mapped into the upper half of virtual memory of every user application (Domain B).
In \cref{sec:sandboxing}, we present a case study with these two security domains.
Another scenario is the execution of SGX enclaves, which share the address space with the host application. 
In \cref{sec:sgx}, we present a case study in which \Approach ensures that a malicious or exploited enclave (Domain B) cannot modify data of the host application (Domain A).

\subsection{Threat Model}\label{sec:threat-model}
For \Approach, we assume two different security domains in one application, such as user and kernel space, or untrusted application and trusted enclave. 
\Approach provides additional sandboxing to existing isolation mechanisms and hence assumes that the used isolation mechanisms are reliable. 
In particular, \Approach presumes that the MMU-based isolation cannot be circumvented architecturally. 
Transient-execution attacks circumventing security domains do not undermine the security of \Approach. 
While Meltdown~\cite{Lipp2018meltdown} showed that the US bit can be circumvented during transient execution, it only allows reading the page. 
Similarly, Foreshadow~\cite{Vanbulck2020lvi} circumvented the present bit for reading enclave memory. 
However, \Approach does not protect the confidentiality. 
Hence, Meltdown or Foreshadow do not cause a security problem. 
Similarly, while Spectre v1.1~\cite{Kiriansky2018speculative}, Store-to-Leak~\cite{Schwarz2019STL}, and LVI~\cite{Vanbulck2020lvi} showed that values can be transiently written to inaccessible pages, applications can simply issue a serializing instruction before reading from the \Approach-isolated memory region to ensure that the architectural value is read. 
Fault attacks~\cite{Kim2014,Murdock2019plundervolt,Kenjar2020v0ltpwn} are out of scope.

\section{Case Studies}\label{sec:case-studies}
In this section, we present two case studies showing that \Approach efficiently isolates different domains.
First, we apply our method to facilitate efficient and complex \syscall argument filtering that is not vulnerable to TOCTOU vulnerabilities (\cref{sec:sandboxing}).
Second, we apply \Approach to SGX, preventing attacks from untrusted or exploited enclaves~\cite{Schwarz2019SGXMalware} (\cref{sec:sgx}).

\subsection{Enhanced \Syscall Filtering}\label{sec:sandboxing}
To restrict \syscalls, Linux provides developers with seccomp.
While seccomp allows filtering static, primitive \syscall arguments, it does not support complex arguments such as strings or structs due to TOCTOU vulnerabilities~\cite{Edge2019,Edge2020}.
Other approaches are based on \syscall interposition, where the \syscall is delegated to another process which decides whether a \syscall is allowed~\cite{Wagner1999,Ford2008,garfinkel2004ostia,Ghormley1998,Provos2003}.
Unfortunately, additionally to the cost of delegating the \syscall, these approaches suffer from the same TOCTOU problem as seccomp~\cite{garfinkel2003traps,Watson2007}.
Hence, this is so far an unsolved problem.

We focus on deep argument filtering without introducing TOCTOU vulnerabilities while maintaining the performance of seccomp in typical scenarios.
By applying \Approach, we remove an attacker's ability to modify the \syscall argument while the kernel performs operations, such as checks, on it. 
Specifically, we prevent the attacker from rewriting a pointer or modifying a string while the kernel inspects that \syscall argument.
On a high level, we identify the page used by the \syscall argument and make it read-only or fully inaccessible by modifying specific bits in the page-table entry.
The argument is then compared to the allowed values for the respective \syscall, and if the argument check succeeds, the \syscall is executed.
By design, this approach cannot suffer from TOCTOU vulnerabilities.
While such an approach sounds straightforward, it requires solving several challenges such as process creation and replacement, multithreading, alias mappings, and other memory-access interfaces.

By providing the means for deep argument filtering, we can improve the security of the system by further limiting the post-exploitation impact of a memory safety vulnerability as we can better restrict \syscalls such as \textit{exec}.
All other seccomp functionality, \ie filtering \syscalls without argument checking and checking of static integer arguments, is naturally supported.
We refer to this as \textit{simple filtering}.

\subsubsection{Threat Model}
Beyond the threat model of \Approach (\cf \Cref{sec:threat-model}), we assume that the application itself is not malicious but exploitable, \eg due to a memory-safety violation, allowing an attacker to gain arbitrary code execution within the application.
Similarly, the kernel is considered to be trusted but potentially exploitable.
We assume that the post-exploitation phase targets the system and requires \syscalls, \eg to gain kernel privileges.
Contrary to previous work~\cite{Ghavamnia2020,Canella2021chestnut,DeMarinis2020}, we can filter \syscalls related to file operations as we can inspect complex data types.
Our approach is orthogonal to other defenses such as CFI, ASLR, NX, or canary-based protections and improves security if these were circumvented.

\subsubsection{Implementation}
Our enhanced sandboxing consists of two parts: a kernel part that performs the actual task of filtering \syscalls and the respective arguments, and a support library that can be linked to the application.
The support library provides functionality for generating the individual \syscall filters and installing them in our kernel module.
As such, our support library is a similar high-level abstraction as \textit{libseccomp} is for seccomp. 
However, filter generation in libseccomp is much more complex due to the usage of BPF.
As our filters are not expressed in BPF, the setup is significantly easier.
For our proof-of-concept implementation, we implement our filtering entirely as a standalone kernel module. 
The kernel module has the advantage that it can be used with any kernel version without recompiling the kernel. 
For a production-ready system, the filtering should be implemented directly in the kernel.

A prerequisite for all \syscall-filtering approaches is to be able to intercept \syscalls. 
Fortunately, entry points for \syscalls are clearly defined, making it easy to intercept all \syscalls.
This includes both legacy 32-bit \syscalls as well as 64-bit \syscalls. 

\begin{listing}[t]
 \begin{lstlisting}[language=C,style=customc]
filter_info_t *filters = dpti_create_filters();
dpti_add_filter_rule(filters, SYS_read);
dpti_add_filter_rule_string(filters, SYS_write, 1, EQ, "teststring");
dpti_install_filters(filters);
\end{lstlisting}
\vspace{-1.5ex}
\caption{Example of a simple allow/reject filter for \textit{read} and a complex filter requiring deep argument filtering on the first argument of the \textit{write} \syscall using our support library.}
\label{lst:filters}
\end{listing}

\paragraph{Setup.}
For the setup, the user-space application sends the filters to the kernel module.
The filters are defined in a high-level representation similar to libseccomp, as illustrated in \Cref{lst:filters}. 
The module then performs a deep copy of the filters into kernel memory.
Preventing TOCTOU vulnerabilities is not necessary at this point, as the application is still considered benign.
Otherwise, if the application was already exploited, an attacker can either manipulate the filters or skip their installation entirely, rendering the filtering useless.
This is in-line with other filtering approaches, such as seccomp-bpf.
Once the filters have been copied, the application is considered sandboxed, and updates to the filters are no longer possible.
A full implementation can consider allowing further restrictions of the filters, similar to seccomp.

\paragraph{Filtering Syscalls.}
Every requested \syscall is delegated to our generic \syscall function inside the module.
If the \syscall originates from a sandboxed application, the generic \syscall function uses the \syscall number to retrieve the \syscall's filter rules.
Contrary to seccomp, checking the filter for a specific \syscall does not require scanning all \syscall filters. 
Instead, it is a simple array access; the lookup time for a filter does not depend on its position within the set of filters~\cite{Hromatka2018}.
This reduces the asymptotic runtime from linear to constant.

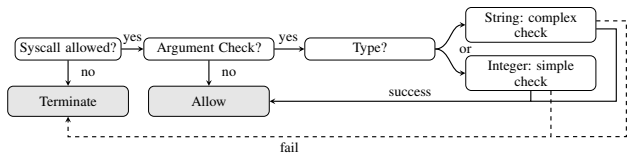
\begin{figure}[t]
  \resizebox{1\hsize}{!}{
    \usetikzlibrary{positioning}

\tikzstyle{startstop} = [rectangle, rounded corners, minimum width=3cm, minimum height=0.75cm,text centered, draw=black,fill=black!2]
\usetikzlibrary{shapes.geometric, arrows}
\tikzstyle{kernel} = [rectangle, minimum width=3cm, minimum height=0.75cm, text centered, draw=black,fill=black!2]
\tikzstyle{process} = [rectangle, rounded corners, minimum width=3cm, minimum height=0.75cm, text centered, draw=black, fill=black!10]
\tikzstyle{decision} = [diamond, minimum width=2cm, minimum height=0cm, text centered, draw=black, fill=white,thick]
\tikzstyle{arrow} = [thick,->,>=stealth]

\begin{tikzpicture}

\node[draw,rounded corners] (first_check) at (0, 0) {Syscall allowed?};

\begin{scope}[shift={(3.5,0)}]
  \node[rectangle,draw,rounded corners] at (0,0) (argument_check) {\parbox{3cm}{\centering Argument Check?}};
\end{scope}

\begin{scope}[shift={(7.5,0)}]
  \node[rectangle,draw,rounded corners] at (0,0) (type) {\parbox{3cm}{\centering Type?}};
\end{scope}

\begin{scope}[shift={(11.5,0)}]
  \node[rectangle,draw,rounded corners] at (0,0.6) (string) {\parbox{3cm}{\centering String: complex check}};
  \node[rectangle,draw,rounded corners] at (0,-0.6) (integer) {\parbox{3cm}{\centering Integer: simple check}};
\end{scope}

\begin{scope}[shift={(0,-1.3)}]
  \node[rectangle,process] at (0,0) (terminate) {Terminate};
  \node[rectangle,process] at (3.5,0)(success) {Allow};
\end{scope}

\draw[arrow] (first_check.south) -- (terminate.north) node[midway,xshift=0.5cm] {no};
\draw[arrow] (first_check.east) -- (argument_check.west) node[midway,above] {yes};
\draw[arrow] (argument_check.south) -- (success.north) node[midway,xshift=0.5cm] {no};
\draw[->,>=stealth,out=0,in=180,thick] (argument_check.east) to (type.west) node[above,xshift=-0.4cm] {yes};
\draw[->,>=stealth,out=0,in=180,thick] (type.east) to (string.west);
\draw[->,>=stealth,out=0,in=180,thick] (type.east) to (integer.west) node[above,yshift=0.375cm] {or};
\draw[arrow] (integer.south) |- (success.east) node[midway,above,xshift=-3cm] {success};
\draw[arrow,dashed] ($(integer.south)+(0.5,0)$) |- ($(terminate.south)+(0,-0.5)$) node[midway,below,xshift=-6.5cm] {fail} -| (terminate.south) ;

\draw[thick,>=stealth] ($(string.east)+(0,-0.1)$) -- ($(string.east)+(0.5,-0.1)$)  |- ($(integer.south)+(0,-0.26)$);
\draw[thick,>=stealth,dashed] ($(string.east)+(0,0.1)$) -- ($(string.east)+(0.75,0.1)$)  |- ($(integer.south)+(0.55,-1.14)$);
\end{tikzpicture}
  }
  \caption{Overview of the 3 \Approach \syscall-filtering cases. \hfill \,}
  \label{fig:filtering-illustration}
\end{figure}

We consider \SIx{3} different cases when checking a \syscall, as illustrated in \cref{fig:filtering-illustration}:

\textit{(1) \Syscall not allowed:}
The \syscall fails the initial check whether it is allowed, making it unnecessary to check potential argument filters.
If a \syscall is executed that was not registered, the application can be terminated. 
Hence, we can perform an early out and kill the process.

\textit{(2) \Syscall allowed, no argument filtering requested:}
The \syscall passes the initial check, \ie the application explicitly allowed the \syscall.
This necessitates the check of potential argument filters.
However, the \syscall was allowed unconditionally as no filters are present.
Therefore, \Approach calls the original \syscall function with the same arguments.
We refer to these first two cases as simple allow and reject filters.

\textit{(3) \Syscall allowed, argument filtering requested:}
The \syscall itself is allowed, making it necessary to check the installed argument filters.
As the ABI~\cite{Matz2014abi} defines that \syscalls can have up to \SIx{6} arguments, every argument must be checked against possible filters.
We iterate over all potential arguments, checking for each whether a filter is defined.
Checks then distinguish between primitive and complex data types.
Primitive data types fully fit into the 64-bit register used as \syscall argument. 
Such data types include, \eg integers, booleans, flags, or single characters. 
Checking these primitive data types does not require any special handling as these arguments are already copied to the kernel.
Hence, they are not vulnerable to TOCTOU and are similarly handled as they are in seccomp.

Complex data types cannot be contained fully in the 64-bit register and are typically pointers to either strings or structures.
These pointers point to user-space memory and can thus be modified by a concurrently running thread during the \syscall.
If the modification happens between applying the filter and executing the \syscall, the filter is effectively rendered useless. 
We focus on checking string arguments, as strings are widely used as \syscall parameters.
As string parameters are especially used with file names~\cite{Schwarz2018DF}, they are an excellent target for sandboxing. 
Our method is not limited to strings but also applies to structures. 
For these types of arguments, we have to ensure that they cannot be modified between applying the filter and the execution of the \syscall.
Thus, we first resolve the page table mappings of the argument pointer.
At this point, we can differentiate between our two variants of \Approach, providing different behavior and security guarantees.

\textit{\ApproachHide:}
Our first variant relies on modifying the US bit, essentially making the page a kernel page and fully inaccessible to user space.
It first attests whether the executable bit is not set in the resolved page-table entry as we do not want to bring user-controlled executable pages into the kernel due to security concerns.
If this bit is set, our implementation kills the application.
Otherwise, the US bit is cleared, \ie flushing the corresponding TLB entry from all necessary CPU cores.
When an attacker tries to modify the argument in a different thread, the MMU enforces the protection by generating a page fault due to the privilege violation.
Our module can now perform a simple string comparison with all developer-provided values without having to consider the possibility of a TOCTOU vulnerability.
If the argument fulfills the \syscall filter rule, the \syscall is allowed, and the original \syscall is executed.
Before returning from the \syscall, arguments are restored to user-space accessible if the physical page is currently not used in another, concurrently executed, \syscall where string filters are used.
Otherwise, the application is killed due to the violation.

\textit{\ApproachFreeze:}
The second variant relies on the RW bit instead of the US bit, making the page write-protected instead of fully inaccessible.
By clearing it in the kernel data structures as well as in the page-table entry and flushing the page from the TLB, the page is read-only.
A malicious thread can only read from it but no longer modify its content.
Thus, TOCTOU vulnerabilities are no longer possible.
The \syscall filtering is equivalent to \ApproachHide. 
After the \syscall is done executing, we again set the previously cleared bits if they were set before the \syscall.
This requires some additional tracking of pages to determine whether this step is necessary.
To prevent a concurrent thread from issuing an \textit{mprotect} \syscall to make the page writable again during the execution of another \syscall, it is necessary to stall \textit{mprotect} with write permissions on a page that had its write permissions cleared by \ApproachFreeze.
The functionality for stalling is already available in the Linux kernel, as it is also necessary, \eg for swapping pages. 
We can leverage the same functionality.

\subsubsection{Special Cases}
The approach outlined above works for most \syscalls and can be handled by our generic \syscall function. 
Only creating (\textit{fork}, \textit{clone}) and destroying (\textit{exit}) processes need additional treatment as well replacing an existing process with \textit{exec}.
Linux creates a new task for a newly created process or thread, including a new process identifier.
Hence, for the \texttt{fork} and \texttt{clone} \syscall, we increase the reference counter of the filters and share them with the newly created process or thread.
When calling \texttt{exec} after \texttt{fork}, we ensure that the copy-on-write \syscall-parameter page is made accessible again for the other process(es) after the \texttt{exec} succeeds. 
In our proof of concept, we simply trigger a copy-on-write fault on this page before manipulating the page-table entry.

A second special case is the handling of the \textit{exit} and \textit{exit\_group} \syscalls, as they are responsible for cleaning up our filters.
When a currently sandboxed application executes one such \syscall, we decrease the reference counter of the filters by the number of threads in the group and mark the thread group as no longer sandboxed.
If the reference counter reaches \SIx{0}, \ie no thread or forked process needs the filters anymore, we free the memory before we terminate the process.

\subsubsection{Multithreading}\label{sec:sandboxing-multithreading}
Multithreading needs special attention.
\Approach works on a page-size granularity. 
Hence, it is likely that the string is not the only content of the associated memory page.
While read accesses from a concurrent thread are no problem for \ApproachFreeze, write accesses to the page lead to a page fault.
For \ApproachHide, both read and write accesses trigger a page fault due to the violation of the privilege boundary.
These accesses might indicate a potential exploitation attempt. 
However, it is more likely that they are part of a legitimate access to data on the page that is unrelated to the \syscall.

To handle such situations, we have to change the page-fault handler slightly. 
In the case of \ApproachHide, the page-fault handler can easily determine that the access would normally be legal, as kernel pages are not found within the user-space address range.
We experimentally verified this by scanning the pages of all running user-space processes on an Ubuntu 18.04 machine.
If the page-fault handler determines that the faulting page is not an actual kernel page, it stalls the offending thread until the \syscall has finished and the page is again available to user space.
The kernel already needs to stall processes for swapping; the same functionality can be used in this case.
We discuss the potential stalling times in more detail in \cref{sec:sandboxing-eval}.

For \ApproachFreeze, we already track all pages that are modified for deep argument filtering, allowing us to easily differentiate a potentially legal write from one to a page that was never writable.
If the page fault occurs on an access to such a tracked page, we again stall the offending thread until the original page access rights are restored.
As stated before, read accesses cause no problem for this variant.

\subsubsection{Alias Mappings}\label{sec:sandboxing-alias-mappings}
Although multiple \textit{writable} mappings to the same physical address are rare (\cf \cref{sec:sandboxing-eval:perf}), \Approach has to consider such scenarios as well. 
If an attacker has code execution, \eg due to a memory safety vulnerability, the attacker can create two virtual address mappings to the same physical page, \ie by using \textit{mmap} and other shared-memory-related \syscalls.
In such a case, the virtual address mapping used by the \syscall is stashed or frozen, but the content of the physical page can be modified via the second, unaltered mapping.
This once again facilitates a TOCTOU vulnerability.
Hence, it is necessary that alias memory mappings are tracked, and if one such mapping is used during the argument check, all other mappings to the same page must be modified as well.

To cover all possible methods of creating an alias mapping, we incorporate our tracking via a probe on the page-fault handler.
This allows us to track alias mappings that are allocated by pre-faulting the page, \ie \textit{mmap} in combination with MAP\_POPULATE, or a page fault upon the first access.
For each physical page, we store all virtual addresses mapping this page, including meta information such as the permission and the process, independent of whether the mapping is in the same or a different process.
This information is then used when checking a string argument during a \syscall by iterating over the alias mappings of the currently used page and modifying the mapping as previously outlined.
We discuss the impact on the performance as well as how frequently such mappings are used by real-world software in \cref{sec:sandboxing-eval:perf}.

\subsubsection{Modifying Memory through /proc/self/mem}
Finally, an attacker can potentially circumvent our TOCTOU-free argument filtering via the \textit{/proc/self/mem} interface, which allows directly modifying the content of a physical page, ignoring missing write permissions.
To protect against this possibility, it is necessary to either restrict the access to this file entirely, \ie make it read-only, or to the offset that corresponds to the physical page currently used in a \syscall.
In our proof-of-concept, we rely on the former by installing a probe on the \textit{mem\_write} function that prevents the write if the application is currently sandboxed.
As the read-only approach is already done by REL 5 and 6, it is reasonable to assume that this does not impair the functionality of applications~\cite{Matousek2016procmem}.
Moreover, to understand whether this interface is used in open-source software, we looked at all code on GitHub that opens this file for writing. 
After filtering out all PoCs for exploits, we only found one project (\texttt{rr}) that uses this functionality. 
Hence, while a production-ready implementation can implement more fine-grained control, it is questionable whether write access to this file is necessary at all.

\begin{tcolorbox}[boxsep=1pt,left=2pt,right=2pt,top=1pt,bottom=1pt,title={Summary}]
  We presented an alternative for seccomp that provides improved performance for simple filters due to less time-consuming checks.
  We additionally provide two variants for deep argument filtering without being vulnerable to TOCTOU vulnerabilities by efficiently and securly modifying page-table entries.
  The latter is currently not supported by seccomp; hence we further improved the systems' security.
\end{tcolorbox}

\subsection{SGX-Protection Domain}\label{sec:sgx}
The classical SGX threat model is asymmetric, \ie it enforces strong protection guarantees for enclaves but does not protect the user application from the loaded enclave.
Enclaves can lead to various security concerns, as neither the operating system nor the user application can verify the code executed inside the enclave, as intended by the design of SGX.
One limitation of SGX is that it does not allow the enclave to execute code outside the enclave boundaries~\cite{Costan2016sgx}.
However, Schwarz \etal~\cite{Schwarz2019SGXMalware} showed malware inside enclaves can simply manipulate the user applications' stack to execute arbitrary code outside the enclave.
To balance the protection guarantees for the user application, SGXJail~\cite{Weiser2019SGXJail} proposed to isolate enclaves by moving them to a separate process with strict \syscall filters, or by relying on hardware modifications, \ie memory protection keys (Intel MPK).

Our \Approach-based SGX protection domain extends the memory isolation guarantees of SGXJail.
However, with \Approach there is no increased overhead of process isolation, no need for MPKs, and no need for changing the SGX specification.
We apply \Approach by replacing a thread's page-table mapping and isolate the user application's memory during the execution of an enclave.
As in this case, most available pages, \ie the entire user-space application, have to be isolated, switching the mapping is faster than iterating over all mapped pages. 

The isolated mapping only contains the original mappings for the loaded enclave, a single additional code page mapping, and a few data page mappings from the user application.
The single code-page mapping is used to enter the enclave while the additional data-page mappings are used to transfer data from and to the enclave.
Upon enclave exit, our protection verifies that the enclave did not tamper with the provided pages or registers and restores the non-isolated mapping for normal execution.
All other pages are either protected by \ApproachFreeze or \ApproachHide, depending on what security guarantees the user application enforces (\cf \Cref{sec:sgx-eval}).

In contrast to our proposed \syscall filtering, the SGX protection domain implements \ApproachHide by not mapping a page instead of clearing the US bit.
Therefore, the enclave can no longer access any of the non-isolated pages as these pages are no longer mapped inside the thread's virtual address space.
When using \ApproachFreeze, all of the non-isolated pages are mapped as read-only pages in the isolated mapping, allowing the enclave to read but not modify them.

To perform fast transitions, we construct the isolated mapping once and then reuse and update it upon entering.
We redesign the data flow between the enclave and the user application to remove unnecessary copying of the data passed to an enclave, leading to a high-performance SGX protection domain with only an overhead of \SIrange{9.9}{24.0}{\percent} for the worst case scenario (\cf \Cref{sec:sgx-eval}).

\subsubsection{Threat Model}
For the SGX protection domain, we reverse the classical SGX threat model and assume that the loaded enclave is untrusted and potentially malicious, whereas the app loading the enclave is trusted.
We assume that the enclave tries to read or modify data of the user application, \eg to mount a ROP attack~\cite{Schwarz2019SGXMalware}.
The trusted user application does not use seccomp filtering to restrict syscalls.
We assume that the code executed inside the enclave is unknown and imposes no restrictions on limiting the interfaces between the enclave and the user application in the sense of ECALLs and OCALLs.

\subsubsection{Implementation}\label{sec:sgx-impl}
The SGX protection domain requires no changes to existing enclaves.
All parts are implemented in the driver and the Untrusted RunTime System (URTS).

\paragraph{Isolated Mapping.}
For the isolated mapping, we extend the SGX driver, which provides the necessary information about pages associated with a given enclave.
We create one isolated mapping per enclave. 
This design decision ensures that multiple threads within enclaves are supported while colluding enclaves cannot circumvent the isolation. 

The enclave's isolated mapping is extended by a Code Bridge Page (CBP) and Data Bridge Pages (DBPs). 
The CBP contains the \texttt{EENTER} instruction used to enter enclaves inside the URTS. 
It is shared amongst all enclaves as the URTS library is only loaded once inside the user application.
The Data Bridge Pages (DBPs) are used to transfer data between the enclave and the user application.
To facilitate fast switches between two mappings, we only exchange the actual hardware mapping inside the \texttt{CR3} register and retain the VMA structures of the original mapping.

\begin{figure}[t]
  \resizebox{0.9\hsize}{!}{
    \definecolor{mycolorlvl1}{RGB}{219, 48, 122}
\definecolor{mycolorlvl2}{RGB}{219, 48, 122}
\definecolor{mycolorlvl3}{RGB}{219, 48, 122}

\begin{tikzpicture}  

\pgfmathsetmacro{\width}{\hsize/3}
\pgfmathsetmacro{\height}{7.0cm}

\pgfmathsetmacro{\firstdist}{0.75}

\pgfmathsetmacro{\yscale}{0.75}
\pgfmathsetmacro{\xscale}{1}

\pgfmathsetmacro{\offset}{0.4}
\pgfmathsetmacro{\ly}{0.1}

\begin{scope}[yscale=\yscale, xscale=\xscale]

\tikzstyle{parent} = [rectangle, draw=black, text width=1.4*\width,rounded corners=5pt,anchor=north]
\tikzstyle{child}  = [rectangle, draw=black, text width=0.6*\width,rounded corners=2pt,anchor=north]

\tikzstyle{childtext} = [rectangle, text width=0.5*\width, minimum height=0.5cm]

\node [parent, minimum height=7cm]  at (-5.2, 1) (app)     {};
\node [childtext, below right] at (app.north west) (app_title) {\textbf{Application}};

\node [] at (-5+\offset, 0.25) (tmp1) {};

\node [child, minimum height=2cm, below=0.0cm of tmp1,pattern=north west lines,pattern color=red!40] {};
\node [childtext, below=0.0cm of tmp1] (syscallh)    {\texttt{\dots}};
\node [childtext, below=0.5cm of tmp1]               {};
\node [childtext, below=1.0cm of tmp1]               {\texttt{}};
\node [childtext, below=1.5cm of tmp1] (syscall)     {\texttt{syscall}};

\node [left=0.2cm of syscallh]           {\small \texttt{0x..1000}};
\node [right=0.1cm of syscall]            {\raisebox{.5pt}{\textcircled{\raisebox{-.9pt} {1}}}};

\node [] at (-5+\offset, 0.25-2.75) (tmp2) {};

\node [child, minimum height=2cm, below=0.0cm of tmp2,pattern=north east lines,pattern color=cyan!40] {};
\node [childtext, below=0.0cm of tmp2] (eenter)        {\texttt{eenter}};
\node [childtext, below=0.5cm of tmp2]                 {\texttt{nop}};
\node [childtext, below=1.0cm of tmp2]                 {\texttt{\dots}};
\node [childtext, below=1.5cm of tmp2] (eenterf)       {\texttt{nop}};

\node [left=0.2cm of eenter]           {\small  \texttt{0x..2000}};
\node [right=0.1cm of eenter]           {\raisebox{.5pt}{\textcircled{\raisebox{-.9pt} {2}}}};

\node [] at (-5+\offset, 0.25-2*2.75) (tmp3) {};

\node [child, minimum height=2cm, below=0.0cm of tmp3,pattern=north west lines,pattern color=red!40] {};
\node [childtext, below=0.0cm of tmp3] (faulth)        {\texttt{cmp RDI}};
\node [childtext, below=0.5cm of tmp3]                 {\texttt{jne}};
\node [childtext, below=1.0cm of tmp3]                 {};
\node [childtext, below=1.5cm of tmp3] ()              {\texttt{\dots}};

\node [left=0.2cm of faulth]           {\small \texttt{0x..3000}};
\node [right=0.1cm of faulth]           {\raisebox{.5pt}{\textcircled{\raisebox{-.9pt} {4}}}};

\node [parent, minimum height=3cm]  at (-0.2, 1) (app)     {};
\node [childtext, below right] at (app.north west) (app_title) {\textbf{Enclave}};

\node [] at (0+\offset, 0.25) (tmp1) {};

\node [child, minimum height=2cm, below=0.0cm of tmp1,pattern=north east lines,pattern color=cyan!40] {};
\node [childtext, below=0.0cm of tmp1] (eexith)    {\texttt{\dots}};
\node [childtext, below=0.5cm of tmp1]               {\texttt{RAX=4}};
\node [childtext, below=1.0cm of tmp1] (eexit)              {\texttt{eexit}};
\node [childtext, below=1.5cm of tmp1]      {\texttt{\dots}};

\node [left=0.2cm of eexith]           {\small \texttt{0x..1000}};
\node [right=0.1cm of eexit]            {\raisebox{.5pt}{\textcircled{\raisebox{-.9pt} {3}}}};

\node [rectangle, draw=black, rounded corners=5pt,anchor=east, pattern=north east lines,pattern color=cyan!40]  at (2, -3.35) ()     {\footnotesize Isolated \texttt{CR3}};
\node [rectangle, draw=black, rounded corners=5pt,anchor=east, pattern=north west lines,pattern color=red!40]  at (2, -3.95) ()     {\footnotesize Non-Isolated \texttt{CR3}};

\node [parent, minimum height=3cm]  at (-0.2, -4.3) (app)     {};
\node [childtext, below right] at (app.north west) (app_title) {\textbf{Kernel}};

\node [] at (0.44+\offset, 0.25-2*2.75) (tmp1) {};
\node [] at (0.5+\offset, 0.25-2*2.75+0.1) (tmp1n) {{\footnotesize PF-Handler} \raisebox{.5pt}{\textcircled{\raisebox{-.9pt} {4}}}};

\node [child, minimum height=2cm, below=0.0cm of tmp1,pattern=north west lines,pattern color=cyan!40] {};
\node [childtext, below right=0.0cm and -1.2cm of tmp1] (eexith)    {\texttt{CR3=\&orig}};
\node [childtext, below=0.5cm of tmp1]             {\texttt{}};
\node [childtext, below=1.0cm of tmp1] (eexit)     {\texttt{}};
\node [childtext, below=1.5cm of tmp1]             {\texttt{\dots}};

\node [] at (-1.6+\offset, 0.25-2*2.75) (tmp1) {};
\node [] at (-1.6+\offset, 0.25-2*2.75+0.1) (tmp1n) {{\footnotesize Syscall} \raisebox{.5pt}{\textcircled{\raisebox{-.9pt} {1}}}};

\node [child, minimum height=2cm, below=0.0cm of tmp1,pattern=north west lines,pattern color=cyan!40] {};
\node [childtext, below right=0.0cm and -1.2cm of tmp1] (eexith)    {\texttt{CR3=\&isol}};
\node [childtext, below=0.5cm and -1.2cm of tmp1]             {\texttt{}};
\node [childtext, below=1.0cm and -1.2cm of tmp1] (eexit)     {\texttt{}};
\node [childtext, below=1.5cm and -1.2cm of tmp1]             {\texttt{\dots}};

\end{scope}

\end{tikzpicture} 
  }
  \caption{Control-flow transition of entering and leaving \Approach-based SGX protection domain.}
  \label{fig:sgx-transitions}
\end{figure}
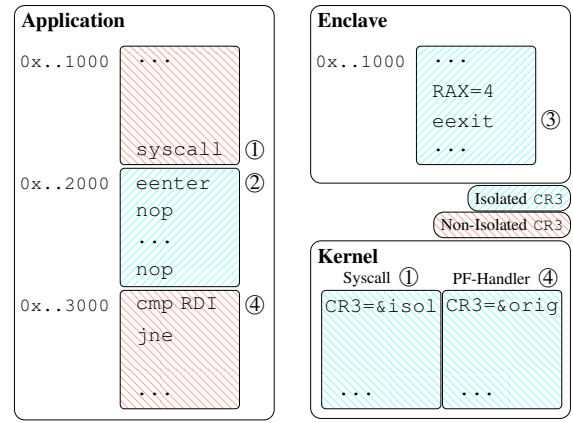

\paragraph{Entering and Exiting Isolation.}
\Cref{fig:sgx-transitions} depicts the control flow for switching protection domains. 
When a thread executes an IOCTL syscall into the SGX driver \textcircled{\raisebox{-.9pt}{1}}, the driver switches the calling thread's virtual address mappings to the isolated address mapping.
\Approach ensures that the \syscall instruction is the last instruction on the page before the CBP. 
Hence, the page containing the \syscall is not part of the isolated mapping, as returning from the \syscall does not require this page. 

The enclave is entered and executed in isolation via the \texttt{EENTER} instruction \textcircled{\raisebox{-.9pt}{2}}.
To exit an enclave, the \texttt{EEXIT} instruction is invoked from inside the enclave, with the RAX register set to \SIx{4} and the return address specified in RBX \textcircled{\raisebox{-.9pt}{3}}.
While \Approach ensures that the only executable mapping is the CBP, an enclave can potentially still return to an arbitrary location inside the CBP.
Hence, the CBP must not contain exploitable code. 
As x86 relies on variable-length opcode and does not enforce instruction alignment, we opted for not placing any functional code on the CBP. 
Instead, we use the CBP as a trampoline by filling it with single-byte NOP instructions
\Approach then leverages the page fault triggered by the first instruction after the CBP \textcircled{\raisebox{-.9pt}{4}}.
The SGX driver catches the page fault and verifies that the fault originates from the first instruction after the CBP.
If this is the case, the driver switches the address space back to the non-isolated mapping.

\paragraph{Interrupts.}
A special case is an asynchronous enclave exit due to an interrupt or fault. 
The CPU stores the enclave state, hands control to the interrupt or fault handler and then resumes the enclave using an \texttt{ENCLU} instruction.
As this instruction has to be mapped, we place it on the CBP.
We verified that a misaligned jump into this instruction cannot be exploited. 

\paragraph{Data Bridges and Stack.}
\Approach must be aware of the pages used for data exchange between the enclave and host application, hence the dedicated DBPs. 
These DBPs are defined when setting up the isolation.
Note that \Approach only works on page granularity.
Therefore, the user must ensure that the data marked as DBPs is aligned and padded to the page boundary to ensure no additional data is exposed.

Before entering the isolation, \Approach aligns the stack pointer to a page boundary, allowing isolation of pages above the thread's stack pointer.
The stack pages below the aligned stack pointer are accessible from within the isolation, as the SGX-SDK uses the user application's stack to pass arguments to OCALLs.
However, these pages do not contain any data the application uses after the enclave returns. 
If already destroyed stack frames were used for sensitive data, they should be manually zeroed before entering the enclave. 
When returning to the user application, \Approach verifies that the stack and the base pointer match the values stored when entering the enclave during the ECALL, ensuring that the enclave did not alter the stack (\cf \Cref{sec:sgx-eval}).

\paragraph{Page Fault Handling.}
Page faults on EPC pages are allowed, as they can be lazily mapped.
Page faults on host-application pages from isolated threads are only allowed on the first instruction of the page after the CBP, and the signal handler. 
Otherwise, the enclave attempted to access data not marked as DBP or jumped to an isolated code page with the \texttt{EEXIT} instruction, leading to a termination of the enclave.
Note that \Approach can distinguish a fault generated inside an enclave from a malicious attempt to execute the signal handler directly.

\subsubsection{Optimizations}\label{sec:sgx:optimized}
We propose an optimization that can be used if the application does not require the \textit{stat} \syscall, or if an untrusted application cannot abuse this \syscall. 
Instead of switching from the SGX domain to the non-isolated domain using a relatively slow page fault (\cf \Cref{sec:sgx-eval:performance,sec:future-work}), we can rely on a \syscall.

While placing a \syscall instruction on the CBP could potentially lead to exploitation of this instruction, we utilize the nature of the \texttt{EEXIT} instruction.
This instruction requires the \texttt{RAX} register to be set to \SIx{4}.
Hence, an attacker can only execute the \textit{stat} \syscall.
This \syscall typically does not allow any control over an application, nor does it allow to leak valuable information.
To communicate with the driver, we hardcode the necessary register values before the \syscall.
Thus, the enclave can only execute the \syscall communicating with the driver or the \textit{stat} \syscall by jumping directly to the \syscall instruction.

\begin{tcolorbox}[boxsep=1pt,left=2pt,right=2pt,top=1pt,bottom=1pt,title={Summary}]
  We presented a solution to confine malicious enclaves, inverting the typical SGX threat model, in a secure way without relying on additional hardware modifications.
  Our solution can be implemented either via \ApproachHide or \ApproachFreeze, each providing different security guarantees.
\end{tcolorbox}

\section{Evaluation}\label{sec:evaluation}
In this section, we evaluate our enhanced \syscall filtering (\Cref{sec:sandboxing}) and our SGX-protection domain (\Cref{sec:sgx}) in terms of security and performance.

\subsection{Enhanced \Syscall Filtering}\label{sec:sandboxing-eval}
We evaluate the performance (\Cref{sec:sandboxing-eval:perf}) and security (\cref{subsec:sandboxing-security}) of both variants of \Approach, clearly showing advantages and disadvantages of \ApproachHide and \ApproachFreeze.

\subsubsection{Performance Evaluation}\label{sec:sandboxing-eval:perf}
In this section, we evaluate the performance of the two variants.
First, we evaluate the performance when executing a single \syscall with reject filters for other \syscalls.
Second, we analyze the overall performance on various real-world applications with simply allow/reject filters.
In both cases, we compare the result to an unsandboxed version and one using seccomp. 
Hence, we only evaluate cases that are supported by seccomp.
Third, we analyze the performance impact of our newly proposed method for deep argument filtering.
Finally, we perform an in-depth analysis of the various steps discussed in \cref{sec:sandboxing}, \ie resolving page tables, bit manipulation, TLB flush, and the string comparison, further substantiating the previous analysis results.

\paragraph{Setup.}
Unless stated otherwise, all experiments in this section are performed on an Intel Skylake i7-6600U running Ubuntu 18.04.1 with kernel version 5.4.0-72-generic at a stable frequency of \SI{2.6}{\giga\hertz}.
To ensure that unrelated mitigations do not affect our measurements, we disabled all mitigations for transient-execution attacks.
For completeness, we verified the functionality of our approach in the presence of them, showing that they do not negatively affect \Approach. 
For our evaluation of seccomp, we rely on the state-of-the-art seccomp library \textit{libseccomp} (2.5.1) to generate and install the respective filters.
Additionally, to improve the performance of seccomp, we enabled the BPF JIT compiler.

\paragraph{Simple \Syscall-filtering Benchmark.}
We first microbenchmark the execution time when filtering a simple \syscall (\textit{getppid}). 
We compare the results to an unsandboxed (vanilla) application and one using seccomp.
We choose \textit{getppid} as it is a fast \syscall without side effects used by previous work~\cite{Canella2021chestnut,Hromatka2018} and by the kernel developers for their benchmarks~\cite{Bueso2019getppid}.
As this experiment does not involve deep argument filtering, and hence no protection of memory contents, there is no difference between \ApproachHide and \ApproachFreeze.

\begin{figure}[t]
  \resizebox{0.9\hsize}{!}{
    \begin{tikzpicture}  
  \begin{axis}[
      ybar, ymin=0,ymax=500, ylabel=Cycles,
      table/col sep=comma,
      legend style={at={(0.5,-0.25)},
	  anchor=north,legend columns=2,draw=none},
      xtick=data,
      xticklabels from table={data/getppid.csv}{approach},
      nodes near coords,
      bar width=0.55cm,
      enlarge x limits ={0.15},
      width=10cm,
      height=4cm,
      ]
      \addplot[fill=green] table [x expr=\coordindex, y=min]{data/getppid.csv};
      \addplot[fill=blue] table [x expr=\coordindex, y=average]{data/getppid.csv};
      \legend{min \qquad, average} 
  \end{axis}

\end{tikzpicture} 
  }
  \caption{\Syscall latency with no, seccomp-, and \Approach-based filtering over \SIx{1} million iterations of the \textit{getppid} \syscall.
  }
  \label{fig:simple-syscall}
\end{figure}
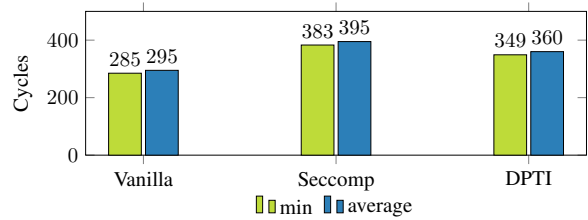

In the sandboxed versions, we allow \SIx{8} and block \SIx{341} \syscalls using simple allow/reject filters.
We measure the average execution time of the \syscall in cycles over \SIx{1000000} executions. 
We show the results of this experiment in \cref{fig:simple-syscall}.

The average execution of a vanilla \textit{getppid} \syscall requires \SIx{295} cycles and \SIx{395} cycles (+\SI{33.9}{\percent}) with seccomp active.
With \Approach, the average execution time is \SIx{360} cycles (+\SI{22}{\percent}), which makes it nearly \SI{10}{\percent} faster than seccomp.
This speed up can be explained by the way the filter check is performed.
While \Approach is constant in the check due to a direct access to the executed \syscalls filter, seccomp either needs to traverse a BPF binary tree or a sequential list of filters.
The least permissive filter for the \syscall then determines whether it is allowed or not.
Other work~\cite{DeMarinis2020} has investigated a \textit{skip-list} to improve the performance of seccomp filters, but as this approach is, to the best of our knowledge, not used in real-world applications we did not use it in our comparison.

\begin{table*}[t]
\centering
\caption{The results of our performance evaluation of unsandboxed, seccomp- and \Approach-sandboxed applications.\
  Overhead shows the percentage overhead compared to our baseline, \ie an unsandboxed version of the respective application.\
  Sample sizes differs for various applications due to the difference in required runtime.}
\label{tbl:isolation-sandboxing-eval}
\resizebox{\hsize}{!}{
\begin{tabular}{llcccc}
& \textbf{Software} & \textbf{Sample Size} & \textbf{\makecell{Normal\\Time / SEM}} & \textbf{\makecell{Seccomp\\Time (Overhead) / SEM}} & \textbf{\makecell{\Approach\\Time (Overhead) / SEM}} \\
\toprule
\multirow{9}{*}{busybox}
 & diff & \SIx{10000} & \SI{0.0025}{\second} / \SIx{7.943e-07} & \SI{0.011}{\second} (\SI{340.0}{\percent}) / \SIx{1.256e-06} & \SI{0.0033}{\second} (\SI{32.0}{\percent}) / \SIx{7.986e-07} \\
 & true & \SIx{10000} & \SI{0.0025}{\second} / \SIx{6.480e-07} & \SI{0.011}{\second} (\SI{340.0}{\percent}) / \SIx{1.324e-06} & \SI{0.0032}{\second} (\SI{28.0}{\percent}) / \SIx{7.263e-07} \\
 & env & \SIx{10000} & \SI{0.0025}{\second} / \SIx{7.044e-07} & \SI{0.011}{\second} (\SI{340.0}{\percent}) / \SIx{1.291e-06} & \SI{0.0032}{\second} (\SI{28.0}{\percent}) / \SIx{7.022e-07} \\
 & ls & \SIx{10000} & \SI{0.0026}{\second} / \SIx{7.202e-07} & \SI{0.011}{\second} (\SI{323.08}{\percent}) / \SIx{1.300e-06} & \SI{0.0033}{\second} (\SI{26.92}{\percent}) / \SIx{6.983e-07} \\
 & dmesg & \SIx{10000} & \SI{0.0025}{\second} / \SIx{8.353e-07} & \SI{0.012}{\second} (\SI{380.0}{\percent}) / \SIx{2.243e-06} & \SI{0.0041}{\second} (\SI{64.0}{\percent}) / \SIx{1.340e-06} \\
 & cat & \SIx{10000} & \SI{0.0025}{\second} / \SIx{8.803e-07} & \SI{0.011}{\second} (\SI{340.0}{\percent}) / \SIx{1.687e-06} & \SI{0.0032}{\second} (\SI{28.0}{\percent}) / \SIx{6.839e-07} \\
 & head & \SIx{10000} & \SI{0.0025}{\second} / \SIx{9.609e-07} & \SI{0.011}{\second} (\SI{340.0}{\percent}) / \SIx{1.410e-06} & \SI{0.0032}{\second} (\SI{28.0}{\percent}) / \SIx{8.144e-07} \\
 & grep & \SIx{10000} & \SI{0.0028}{\second} / \SIx{9.607e-07} & \SI{0.0113}{\second} (\SI{303.57}{\percent}) / \SIx{1.501e-06} & \SI{0.0035}{\second} (\SI{25.0}{\percent}) / \SIx{8.813e-07} \\
 & pwd & \SIx{10000} & \SI{0.0025}{\second} / \SIx{7.664e-07} & \SI{0.011}{\second} (\SI{340.0}{\percent}) / \SIx{1.333e-06} & \SI{0.0032}{\second} (\SI{28.0}{\percent}) / \SIx{8.671e-07} \\
\midrule
\multirow{2}{*}{git}
 & diff & \SIx{10000} & \SI{0.0092}{\second} / \SIx{2.498e-06} & \SI{0.0137}{\second} (\SI{48.91}{\percent}) / \SIx{3.129e-06} & \SI{0.0116}{\second} (\SI{26.09}{\percent}) / \SIx{3.306e-06} \\
 & status & \SIx{10000} & \SI{0.0068}{\second} / \SIx{1.019e-06} & \SI{0.0112}{\second} (\SI{64.71}{\percent}) / \SIx{1.248e-06} & \SI{0.0094}{\second} (\SI{38.24}{\percent}) / \SIx{1.361e-06} \\
\midrule
\multirow{6}{*}{ffmpeg}
 & extract & \SIx{10} & \SI{12.1098}{\second} / \SIx{8.482e-03} & \SI{12.5605}{\second} (\SI{3.72}{\percent}) / \SIx{8.249e-03} & \SI{12.156}{\second} (\SI{0.38}{\percent}) / \SIx{9.288e-03} \\
 & convert & \SIx{10} & \SI{18.6844}{\second} / \SIx{9.543e-03} & \SI{18.9921}{\second} (\SI{1.65}{\percent}) / \SIx{6.754e-03} & \SI{18.8285}{\second} (\SI{0.77}{\percent}) / \SIx{9.704e-03} \\
 & remove & \SIx{10} & \SI{18.2757}{\second} / \SIx{6.121e-03} & \SI{18.4657}{\second} (\SI{1.04}{\percent}) / \SIx{7.092e-03} & \SI{18.4038}{\second} (\SI{0.7}{\percent}) / \SIx{3.423e-03} \\
 & crop & \SIx{10} & \SI{12.3901}{\second} / \SIx{1.467e-02} & \SI{12.4444}{\second} (\SI{0.44}{\percent}) / \SIx{2.614e-02} & \SI{12.4219}{\second} (\SI{0.26}{\percent}) / \SIx{4.593e-03} \\
 & info & \SIx{10000} & \SI{0.0097}{\second} / \SIx{9.124e-07} & \SI{0.0567}{\second} (\SI{484.54}{\percent}) / \SIx{2.150e-06} & \SI{0.0551}{\second} (\SI{468.04}{\percent}) / \SIx{3.205e-06} \\
 & change & \SIx{10} & \SI{18.2882}{\second} / \SIx{9.466e-03} & \SI{18.5225}{\second} (\SI{1.28}{\percent}) / \SIx{9.164e-03} & \SI{18.4518}{\second} (\SI{0.89}{\percent}) / \SIx{1.333e-02} \\
\bottomrule
\end{tabular}
}
\end{table*}

\paragraph{Real-world Application Benchmark.}
While microbenchmarks show the specific effects of \Approach on \syscalls, they do not allow reasoning about real-world application performance impact. 
Again, as no deep argument filtering is involved, there is no difference between \ApproachHide and \ApproachFreeze.

For this evaluation, we need to determine the application's \syscalls.
Fortunately, several automated approaches have been demonstrated to get this information~\cite{DeMarinis2020,Ghavamnia2020,Canella2021chestnut}, which we extend to support \Approach.
Our baseline is a vanilla version of the respective application.
Both seccomp and \Approach are configured to only log \syscall violations to ensure that \syscalls not identified by the automated approach do not crash the tested application.
We show the result of our evaluation in \cref{tbl:isolation-sandboxing-eval}.
The commands used for each application are provided in \cref{tbl:real-world-commands} in \cref{app:commands}.

We first consider the results of small applications and simple tasks of large ones, \ie busbox, git, the ffmpeg info command.
For these applications, the overhead appears huge, ranging between \SI{25}{\percent} and \SI{468}{\percent} for \Approach.
Still, for all cases, the overhead is lower than for seccomp where we measured an overhead between \SI{48}{\percent} and \SI{485}{\percent}.
For more complex tasks in ffmpeg, the overhead never exceeds \SI{0.9}{\percent} with \Approach, while in one case it almost reaches \SI{4}{\percent} for seccomp.

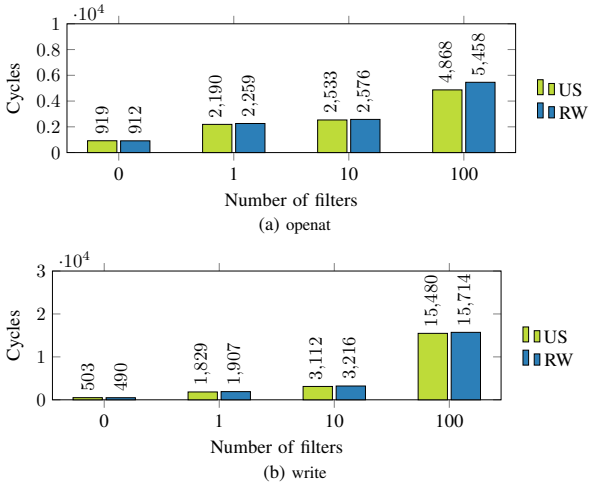
\begin{figure}[t]
  \resizebox{0.9\hsize}{!}{
  \begin{subfigure}[b]{0.6\textwidth}
    \begin{tikzpicture}  
  \begin{axis}[
      height=4cm,
      width=10cm,
      ybar, ymin=0, ylabel=Cycles, ymax=10000,
      every axis plot post/.style={/pgf/number format/fixed},
      table/col sep=comma,
      legend style={at={(1.11,0.6)},
      anchor=north,legend columns=1,draw=none},
      xlabel=Number of filters,
      xtick=data,
      xticklabels from table={data/open_output_no_shared_us.csv}{num filters},
      every node near coord/.append style={rotate=90, anchor=west},
      bar width=0.55cm,
      enlarge x limits ={0.15},
      visualization depends on=rawy\as\rawy, %
      nodes near coords={%
              \pgfmathprintnumber{\rawy}%
          },
      clip=false,
      ]

      \addplot[fill=green] table [x expr=\coordindex, y=average]{data/open_output_no_shared_us.csv};
      \addplot[fill=blue] table [x expr=\coordindex, y=average]{data/open_output_no_shared_rw.csv};
      \legend{US \qquad, RW}
  \end{axis}

\end{tikzpicture} 
    \vspace{-0.5cm}
    \caption{openat}
    \label{fig:string-check-openat}
  \end{subfigure}
  }
  \resizebox{0.9\hsize}{!}{
  \begin{subfigure}[b]{0.6\textwidth}
    \begin{tikzpicture}  
  \begin{axis}[
      height=4cm,
      width=10cm,
      ybar, ymin=0, ylabel=Cycles, ymax=30000,
      every axis plot post/.style={/pgf/number format/fixed},
      table/col sep=comma,
      legend style={at={(1.11,0.6)},
      anchor=north,legend columns=1,draw=none},
      xlabel=Number of filters,
      xtick=data,
      xticklabels from table={data/write_output_no_shared_us.csv}{num filters},
      every node near coord/.append style={rotate=90, anchor=west},
      bar width=0.55cm,
      enlarge x limits ={0.15},
      restrict y to domain*=0:46000, %
      visualization depends on=rawy\as\rawy, %
      nodes near coords={%
              \pgfmathprintnumber{\rawy}%
          },
      clip=false,
      ]

      \addplot[fill=green] table [x expr=\coordindex, y=average]{data/write_output_no_shared_us.csv}; 
      \addplot[fill=blue] table [x expr=\coordindex, y=average]{data/write_output_no_shared_rw.csv};
      \legend{US \qquad, RW} 
  \end{axis}

\end{tikzpicture} 
    \vspace{-0.5cm}
    \caption{write}
    \label{fig:string-check-write}
  \end{subfigure}
  }
  \caption{Cycles required for string filtering for \textit{openat} and \textit{write} \syscalls using \Approach, either using the US or RW bit.
  The x-axis indicates the position of the allowed string within the filter for the \syscall, \ie the number of strings that need to be checked until a match occurs with the current argument.}
  \label{fig:string-check}
\end{figure}

\paragraph{Deep Argument Filtering.}
To evaluate the performance of \Approach for memory isolation, we use both variants for deep argument filtering.
We consider two simple \syscalls, \textit{openat} and \textit{write}.
For both \syscalls, we measure the average execution time of the \syscall, including our deep argument filtering, over \SIx{100,000} invocations.
As a \syscall can allow multiple strings, we consider different number of strings in the filter, placing the correct one at its end, \ie for \textit{n} strings in the filter we require \textit{n} comparisons.
As a baseline, we measure the execution time of the respective \syscall without any argument checking.

\cref{fig:string-check} shows the results of this experiment.
Deep argument filtering has a non-negligible performance overhead:
With \ApproachHide, a single string check increases the execution time of the \textit{openat} \syscall from \SIx{920} to \SIx{2038} cycles. 
This increase is mostly due to the one-time overhead of isolating the memory region. 
With 10 string checks, the execution time is only slightly higher with \SIx{2351} cycles. 
In all cases, the performance of \ApproachHide and \ApproachFreeze is about the same.

\paragraph{In-Depth Analysis of Filtering Components.}
To analyze the overhead further, we perform an in-depth analysis to determine the overhead of each individual step, \ie resolving page tables, bit manipulation, TLB flush, and the actual string comparison.

\begin{table}[t]
  \centering
  \caption{Cycle amounts of the individual components of our deep argument filtering over \SIx{100\,000} executions of the respective \syscall.
    Outliers were replaced with the median.
  }
  \label{tbl:in-depth-analysis}
  \resizebox{0.9\hsize}{!}{
    \begin{tabular}{lllcccc}
      \textbf{\Syscall}       &                     &        & \textbf{Resolving PT} & \textbf{PT Manipulation} & \textbf{Flushing PT} & \textbf{Check} \\
      \toprule
      \multirow{4}{*}{openat} & \multirow{2}{*}{\ApproachHide} & cycles & \SIx{86}                              & \SIx{27}                                 & \SIx{458}                            & \SIx{40}                       \\
                              &                     & SEM    & \SIx{0.0886}                          & \SIx{0.0028}                             & \SIx{0.0533}                         & \SIx{0.0024}                   \\
      \cmidrule{2-7}
                              & \multirow{2}{*}{\ApproachFreeze} & cycles & \SIx{81}                              & \SIx{28}                                 & \SIx{473}                            & \SIx{46}                       \\
                              &                     & SEM    & \SIx{0.0939}                          & \SIx{0.0041}                             & \SIx{0.0683}                         & \SIx{0.0028}                   \\
      \midrule
      \multirow{4}{*}{write}  & \multirow{2}{*}{\ApproachHide} & cycles & \SIx{45}                              & \SIx{27}                                 & \SIx{396}                            & \SIx{137}                      \\
                              &                     & SEM    & \SIx{0.1751}                          & \SIx{0.0028}                             & \SIx{0.0442}                         & \SIx{0.0442}                   \\
      \cmidrule{2-7}
                              & \multirow{2}{*}{\ApproachFreeze} & cycles & \SIx{49}                              & \SIx{27}                                 & \SIx{422}                            & \SIx{185}                      \\
                              &                     & SEM    & \SIx{0.2662}                          & \SIx{0.0034}                             & \SIx{0.1067}                         & \SIx{0.1480}                   \\
      \bottomrule
    \end{tabular}
  }
\end{table}

We instrument \ApproachHide and \ApproachFreeze to measure the required cycles for each step. 
We use the same \syscalls as in the previous experiment, but reduce the number of invocations to \SIx{100\,000}.
We filter outliers---detected using the modified z-score---by replacing them with the median.

The results of the evaluation are shown in \cref{tbl:in-depth-analysis}.
The largest overhead is introduced by the page-table flush, which is necessary to perform twice.
Recent work in Linux 5.13 has improved the performance of TLB flushing~\cite{Larabel2021tlb}, which automatically improves the performance of our filtering as well.
We discuss a potential way to further reduce this overhead in future work in \cref{sec:future-work}.
The string check differs between the two evaluated \syscalls due to different string lengths used in the evaluation.
Interestingly, clearing the additional bit in the kernel data structures required for \ApproachFreeze has no impact.

The execution time of all steps roughly sums up to the time difference shown for the case with no and one string filter in \cref{fig:string-check}.
More efficient caching of page table translations can further improve the performance but was not implemented in our proof-of-concept implementation.

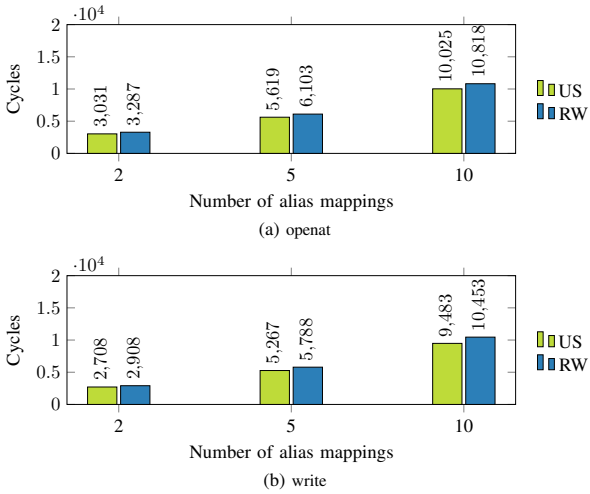
\begin{figure}[t]
  \resizebox{0.9\hsize}{!}{
  \begin{subfigure}[b]{0.6\textwidth}
    \begin{tikzpicture}  
  \begin{axis}[
      height=4cm,
      width=10cm,
      ybar, ymin=0, ylabel=Cycles, ymax=20000,
      every axis plot post/.style={/pgf/number format/fixed},
      table/col sep=comma,
      legend style={at={(1.11,0.6)},
      anchor=north,legend columns=1,draw=none},
      xlabel=Number of alias mappings,
      xtick=data,
      xticklabels from table={data/open_output_shared_us.csv}{num_shared_mappings},
      every node near coord/.append style={rotate=90, anchor=west},
      bar width=0.55cm,
      enlarge x limits ={0.15},
      visualization depends on=rawy\as\rawy, %
      nodes near coords={%
              \pgfmathprintnumber{\rawy}%
          },
      clip=false,
      ]

      \addplot[fill=green] table [x expr=\coordindex, y=average]{data/open_output_shared_us.csv};
      \addplot[fill=blue] table [x expr=\coordindex, y=average]{data/open_output_shared_rw.csv};
      \legend{US \qquad, RW}
  \end{axis}

\end{tikzpicture} 
    \vspace{-0.5cm}
    \caption{openat}
    \label{fig:string-check-openat-alias}
  \end{subfigure}
  }
  \resizebox{0.9\hsize}{!}{
  \begin{subfigure}[b]{0.6\textwidth}
    \begin{tikzpicture}  
  \begin{axis}[
      height=4cm,
      width=10cm,
      ybar, ymin=0, ylabel=Cycles, ymax=20000,
      every axis plot post/.style={/pgf/number format/fixed},
      table/col sep=comma,
      legend style={at={(1.11,0.6)},
      anchor=north,legend columns=1,draw=none},
      xlabel=Number of alias mappings,
      xtick=data,
      xticklabels from table={data/write_output_shared_us.csv}{num_shared_mappings},
      every node near coord/.append style={rotate=90, anchor=west},
      bar width=0.55cm,
      enlarge x limits ={0.15},
      restrict y to domain*=0:46000, %
      visualization depends on=rawy\as\rawy, %
      nodes near coords={%
              \pgfmathprintnumber{\rawy}%
          },
      clip=false,
      ]

      \addplot[fill=green] table [x expr=\coordindex, y=average]{data/write_output_shared_us.csv}; 
      \addplot[fill=blue] table [x expr=\coordindex, y=average]{data/write_output_shared_rw.csv};
      \legend{US \qquad, RW} 
  \end{axis}

\end{tikzpicture} 
    \vspace{-0.5cm}
    \caption{write}
    \label{fig:string-check-write-alias}
  \end{subfigure}
  }
  \caption{Cycles required for string filtering with alias mappings for \textit{openat} and \textit{write} \syscalls using \Approach.
  The x-axis indicates the number of alias mappings for the filtered \syscall argument, including the mapping used in the \syscall itself.}
  \label{fig:string-check-alias}
\end{figure}

\paragraph{Alias Mappings.}
To evaluate the overhead of alias mappings, we build on our previous benchmark with the \textit{openat} and \textit{write} \syscalls.
Each \syscall uses a single string filter, and we vary the number of alias mappings to the \syscall argument, \ie 2, 5, and 10 mappings, including the mapping used in the \syscall.
Each \syscall is executed \SIx{100\,000} and the average execution time is taken.
\cref{fig:string-check-alias} shows the result of this evaluation.

Unsurprisingly, the average execution time of a \syscall increases with the number of alias mappings that have to be modified.
In combination with the results of the previous in-depth analysis (\cref{tbl:in-depth-analysis}) and our performance analysis of non-alias mappings in \syscalls (\cref{fig:string-check}), it is clear that most of the overhead is due to the additional, necessary TLB flushes.
As previously mentioned does recent work on improving the performance of TLB flushes automatically improve the performance of our filtering as well~\cite{Larabel2021tlb}.
Note that this overhead only materializes in the case of at least 2 alias mappings to the same page; otherwise the overhead shown in \cref{fig:string-check} applies.

We also investigate the performance of our tracking itself, which shows that adding a new mapping on average adds \SIx{1222} cycles (N=\SIx{100\,000}, +\SI{47.5}{\percent}) to the page fault.
Most of this overhead, though, is not due to the tracking itself but a limitation of our proof-of-concept implementation as we need to probe (kretprobe) the page-fault handler.
A full implementation can perform the tracking by extending it directly, removing on average \SIx{638} cycles of overhead from the page fault.
Hence, the overhead of the tracking itself is \SI{24.8}{\percent}.
Additionally, a full implementation can modify existing kernel data structures, allowing for a more efficient tracking mechanism.

As we observe a more significant overhead when an alias mapping is used in a \syscall, we now investigate whether such mappings are frequently used in \syscalls in real-world applications.
For this evaluation, we rely on the applications shown in \cref{tbl:isolation-sandboxing-eval}, \ie git, busybox, and ffmpeg, as well as visudo (\cref{sec:evaluation-visudo}).
We track the number of alias mappings created over the execution of each one of our commands and whether such a mapping is then used in a string filter.
This analysis revealed that not a single alias mapping is created and hence not used in a string filter.
Therefore, the additional overhead is non-existent for at least these applications.
We leave a broader analysis of alias mappings in real-world applications for future work.

\subsubsection{Security Evaluation}\label{subsec:sandboxing-security}
For the security evaluation, we first evaluate whether clearing the US- or the RW-bit prevents modification of data.
Then, we demonstrate that we can perform the deep argument check without interference from user space, \ie that our approach does not suffer from TOCTOU vulnerabilities.
Finally, we discuss the security implications of bringing user-controlled pages into the kernel.

\paragraph{Modifying Bits in Page Tables.}
To ensure that the MMU-based isolation is reliable, we create a simple program that allocates a page of memory and then accesses it, observing, as expected, no crash.
In the case of \ApproachHide, we then use PTEditor~\cite{SchwarzPteditor} to clear the US-bit and access it again.
As expected, the second access now results in a crash due to a privilege boundary violation.
For \ApproachFreeze, we modified PTEditor such that it clears the VM\_WRITE bit in the vm\_flags of the associated vm\_area\_struct.
While reading from the page still works, storing data to the page results in a crash due to the violation of the write protection.
Hence, clearing the respective bits prevents another thread from modifying the data once the protection bits have been set appropriately.

\paragraph{Eliminating TOCTOU.}
While the previous experiment shows that the underlying principle works, we verify that it can prevent the exploitation of TOCTOU vulnerabilities.
In this experiment, we try to re-direct an \textit{openat} \syscall such that a wrong file is opened.
We create a multi-threaded application that uses filters that only allow opening the file \textit{file1}.
Thread 1 tries to open the specified file and print its content.
Simultaneously, Thread 2 tries to modify the filename while the \syscall is executed.
In a vanilla or seccomp-based version, the \syscall continues and opens the wrong file.
With \Approach, the page containing the \syscall is unmodifiable for the user.
Hence, Thread 2 triggers a segfault due to the illegal access, and the thread is stalled, mitigating the attack.

\paragraph{User-space Pages in Kernel Space.}
When using \ApproachHide, our enhanced filtering effectively brings a user-controlled page into the kernel space (\cf \cref{sec:sandboxing}).
To ensure that this cannot be exploited to inject arbitrary code into the kernel, \ApproachHide first verifies that the page is not marked as executable. 
As \ApproachHide already has to modify the page-table entry for isolating the page, it can additionally clear the executable bit. 
Hence, as long as the page resides in the kernel space, it is not executable anymore. 
A thread cannot access the page while it is isolated, hence, removing the executable permission does not require any additional handling. 
After the \syscall is done, \ApproachHide can simply restore the executable bit. 
Thus, an attacker cannot exploit \ApproachHide to inject executable pages into the kernel. 
Note that the kernel has to deactivate SMAP to access the \syscall parameters in any case, enabling the access to the entire user-space memory during this time. 
Hence, temporarily bringing one data page into the kernel during a \syscall does not increase the attack surface. 

We consider the possibility of \ApproachHide potentially weakening or breaking KASLR due to the user-space page being brought into the kernel.
As it is only a data page, it can only be used as a stack page in a code-reuse attack where the attacker places the return addresses of the ROP gadgets.
This already requires a KASLR break as the \textit{ret} instruction can only return to absolute addresses.
Hence, we can conclude that \ApproachHide does not weaken or break KASLR.

\subsubsection{Visudo} \label{sec:evaluation-visudo}
To demonstrate how deep argument filtering can improve the security of the system, we harden the \textit{visudo} application with \Approach. 
\textit{visudo} is used to edit the \textit{sudoers} file with automatic validity checks.
It uses an editor from a pre-defined set of editors to edit the file (potentially any editor).
It can also be used to open other files than the \textit{sudoers} file by using a command-line switch.
Interestingly, the tool's manpage already mentions that this behavior can be a security hole:

\emph{
  ``if visudo is configured with the -\@-with-env-editor option or the env\_editor Default variable is set in sudoers, visudo will use any the editor defines by VISUAL or EDITOR. [..] this can be a security hole since it allows the user to execute any program they wish simply by setting VISUAL or EDITOR.''
}

To demonstrate that our enhanced filtering prevents exactly this scenario, \ie an arbitrary editor opening arbitrary files, we manually extend \textit{visudo} with \syscall filters.
We restrict \textit{visudo} to only allow the \textit{vi} editor to open the \textit{sudoers} files as well as files necessary for \textit{vi} to work, \ie configuration files, libraries, and locales.
Hence, in addition to allowing \syscalls required for \textit{visudo}, we also need to allow \syscalls that \textit{vi} requires.

In total, we generate \SIx{52} simple \syscall filter rules without any argument filters.
We allow the \textit{execve} \syscall with a deep argument filter on \textit{/usr/bin/vi}, hence an invocation with a different editor is prevented.
Naturally, the list of pre-defined editors can be extended, and the restriction on \textit{vi} is simply done to ease our proof-of-concept implementation.
We add a deep argument filter on the \textit{openat} \syscall such that it only allows opening the \textit{sudoers} file and all strictly necessary library and file dependencies of \textit{visudo} and \textit{vi}.
This results in \SIx{47} deep argument filters for \textit{openat}.
In total, we define \SIx{100} filters.
While all the simple \syscall filters are supported by seccomp, the complex filters based on string comparisons are not supported. 
Hence, seccomp cannot restrict access to files and editors.

To demonstrate that the resulting \textit{visudo} binary is still able to perform its job, we use it to modify the \textit{sudoers} file using \textit{vi}.
When doing that, we did not observe a single crash, and the task completes successfully.
On the other hand, when we try to override the editor or try to manipulate another file, our module detects that the specified filter rules are being violated and kills the application.
Hence, \Approach can be used to close the previously mentioned security hole.

We evaluate the performance of our protected version of \textit{visudo} over \SIx{10000} executions and compare it to a vanilla version.
For the evaluation, we measure the performance of the non-interactive verification mode of \textit{visudo}. 
Nevertheless, the filtering is the same as in the interactive modification mode, \ie filters for the \textit{openat} \syscall are still checked.
The vanilla version requires, on average, \SI{12.8}{\milli\second}, \ApproachHide \SI{13.2}{\milli\second}, and \ApproachFreeze \SI{13.0}{\milli\second}.

\subsubsection{Comparison \ApproachHide and \ApproachFreeze}
While both variants of \Approach solve the problem of deep argument filtering, they differ in what MMU mechanism enforces the protection.
As the used bits differ in their semantics, the two variants exhibit different properties, which we discuss in more detail now.
The results of our comparison are presented in \cref{tbl:sandboxing-comparison}.

\begin{table}[t]
  \centering
  \caption{Comparison of the two variants of \Approach.
    Write and read possible only consider the ability of the user-space application to read or write, not the kernel.}
  \label{tbl:sandboxing-comparison}
  \resizebox{0.9\hsize}{!}{
    \begin{tabular}{l cccc}
      \textbf{Approach} & \textbf{Modified Bit} & \textbf{Read possible} & \textbf{Write Possible} & \textbf{Tracking}          \\
      \toprule
      \ApproachHide       & US                  & \xmark                                 & \xmark                                  & \xmark \\
      \midrule
      \ApproachFreeze       & \makecell{RW\\VM\_WRITE}                        & \cmark                              & \xmark & \cmark\\
      \bottomrule
    \end{tabular}
  }
\end{table}

\ApproachFreeze requires manipulation of bits in two different locations, but has the advantage that a user-space thread can still read data from the modified page, which is not possible with \ApproachHide.
For \ApproachFreeze, we need to explicitly track the previous permissions of the modified page such that we can restore the correct state after the \syscall completes.
This is not necessary for \ApproachHide.
However, the additional bit to modify does not introduce any overhead (\cref{tbl:in-depth-analysis}).

In both cases, writes are not possible from user space.
Additionally, \ApproachFreeze blocks writes from the kernel.
While this does not provide security benefits, it can be disadvantageous if the kernel wants to write to the isolated page. 
However, we have not encountered such a case in our tests.

\begin{tcolorbox}[boxsep=1pt,left=2pt,right=2pt,top=1pt,bottom=1pt,title={Summary}]
  We presented a detailed performance and security evaluation, showing that \Approach outperforms seccomp in the case of simple \syscall filters while providing additional security functionality in the form of deep argument filtering.
  We highlighted the bottlenecks of the approach and discussed the main differences between \ApproachHide and \ApproachFreeze.
  Finally, we showed on the example of \textit{visudo} how our approach can prevent a real-world security risk.
\end{tcolorbox}

\subsection{SGX-Protection Domain}\label{sec:sgx-eval}
We evaluate the performance (\Cref{sec:sgx-eval:performance}) and the security guarantees (\Cref{sec:sgx-eval:security}) of the SGX protection domain, including the more optimized version from \Cref{sec:sgx:optimized}.

\subsubsection{Performance}\label{sec:sgx-eval:performance}
As page table mappings are enforced by hardware, no additional performance overhead is observed during the regular execution of an enclave.
The protection domain alters the page tables during enclave enter and exit, affecting the latency of enclave transitions only.
We evaluate the overhead using \emph{sgxbench}~\cite{sgxbench}.
For our evaluation, we restrict ourselves to the \emph{empty ECALL} and \emph{empty OCALL} benchmark of \emph{sgxbench} as these model the worst case scenario for our protection.
These benchmarks simply perform ECALLs or OCALLs with no additional code inside the function.

\paragraph{Setup.}
We perform our benchmarks on an \emph{Intel Core i5-8265U} CPU with a fixed frequency of \SI{2.9}{\giga\hertz}.
The operating system is Ubuntu 20.04 with Linux kernel 5.8.0.
We compare \ApproachHide to a reference URTS SGX-PSW library as baseline, and execute each benchmark \SIx{2000000} times.
We also evaluate the performance improvements of the slightly less secure, optimized version (\cf \cref{sec:sgx:optimized}).

\begin{figure}[t]
  \resizebox{1\hsize}{!}{
    \begin{tikzpicture}  
  \begin{axis}[
      ybar, ymin=0,ymax=36000, ylabel=Cycles,
      table/col sep=comma,
      legend style={at={(0.5,-0.35)},
	  anchor=north,legend columns=4,draw=none},
      xtick=data,
      xticklabels={ECALL,OCALL},
      nodes near coords,
      every node near coord/.append style={rotate=45, anchor=west},
      bar width=0.55cm,
      enlarge x limits ={0.75},
      width=10cm,
      height=3.5cm,
      ]
      \addplot[fill=green] table [col sep=semicolon,x expr=\coordindex, y=mean]{data/sgx_ref.csv};
      \addplot[fill=blue] table [col sep=semicolon,x expr=\coordindex, y=mean]{data/sgx_isol.csv};
      \addplot[fill=yellow!50!white] table [col sep=semicolon,x expr=\coordindex, y=mean]{data/sgx_isol_syscall.csv};
      \legend{non-isolated \qquad, page fault \qquad, syscall}  
  \end{axis}

\end{tikzpicture} 
  }
  \caption{Microbenchmark results for ECALLs and OCALLs for a non-isolated enclave, one using a page fault for ending isolation, and one using a \syscall instead of the page fault.}
  \label{fig:sgx-perf}
\end{figure}
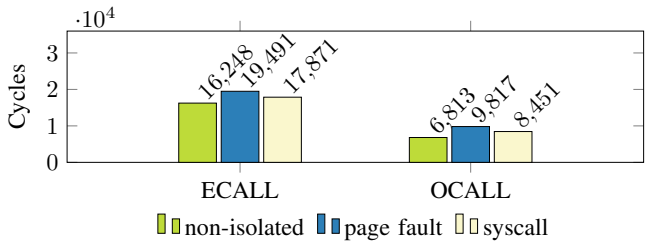

\paragraph{Results.}
\Cref{fig:sgx-perf} shows the benchmark results.
We observe a latency increase of \SI{19.9}{\percent} for ECALLs and \SI{44.0}{\percent} for OCALLs.
When using the optimized version using the \syscall instruction, we observe only an overhead of \SI{9.9}{\percent} for ECALLs and \SI{24.0}{\percent} for OCALLs.
This improvement is solely caused by the different execution times of the page-fault handler and the \syscall handler. 
We discuss how future work can optimize our implementation in \Cref{sec:future-work}.
In similar microbenchmarks, SGXJail~\cite{Weiser2019SGXJail} had an overhead of \SI{41.4}{\percent} for ECALLs and \SI{45.2}{\percent} for OCALLs.
Hence, as shown by the microbenchmarks, optimized \Approach outperforms SGXJail by \SI{22}{\percent} for ECALLs and \SI{14.6}{\percent} for OCALLS.

\subsubsection{Security}\label{sec:sgx-eval:security}
In this section, we evaluate the security the \ApproachFreeze- and \ApproachHide-based enclave isolation.

\paragraph{Data-Only Attacks.}
A non-isolated enclave can read or modify host application memory, hence allowing a malicious enclave to perform data-only attacks.
\Approach prohibits data-only attacks by limiting write access to host pages, allowing the enclave to modify pages marked explicitly as DBPs only.
As these pages are, by definition, used to pass data from and to a possibly untrusted enclave, we require the host to verify the correctness of the data received from the enclave.

\paragraph{SGX-ROP.}
To perform ROP-based attacks, the enclave needs to scan the host application for gadgets to build a ROP chain and then overwrite the stored return address on the stack to execute the ROP chain~\cite{Schwarz2019SGXMalware}.
With both \ApproachFreeze and \ApproachHide, we prohibit modifications to the return address on the stack by aligning the stack downwards before entering isolation and restricting access to the upper stack pages containing the return address.
As the enclave can freely modify stack and base pointer upon leaving isolation, \Approach verifies that the latter was not altered.
For data pushed on the stack, we have to consider the two reasons a thread can exit an enclave:
First, the thread can exit the enclave to return from an ECALL.
In this case, we verify that the stack pointer was restored correctly and restore the saved registers before executing the return instruction.
Second, the enclave exited to perform an OCALL.
The data is pushed onto the user stack, and the protection only verifies that the stack pointer grew downwards.
No return instruction is executed directly after the isolation end as an OCALL adds additional call frames instead of removing them.

\paragraph{EEXIT Destination.}
The \texttt{EEXIT} instruction allows arbitrary return locations outside the SGX enclave, allowing returns to arbitrary, potentially missaligned, instructions~\cite{Weiser2019SGXJail}.
We consider two cases:
First, the enclave returns to a page different than the CBP, raising a page fault.
As discussed in \cref{sec:sgx}, this results in our page-fault handler raising a segmentation fault.
Second, the enclave returns to the CBP, potentially misaligned.
This poses no problem as the CBP is only filled with \texttt{NOP} instructions following the \texttt{EENTER}.

\paragraph{Data Confidentiality.}
\ApproachHide provides data confidentiality as host application pages are not mapped in the isolated address space.
\ApproachFreeze does not guarantee data confidentiality as pages are just mapped as read-only, allowing the enclave to read host data, but it still prevents malicious modification.

\paragraph{Multithreading and TOCTOU.}
As described in \Cref{sec:sgx}, multiple threads inside the same enclave share an isolated mapping.
This does not pose a security problem, as threads within an enclave already share the enclave memory. 
Nevertheless, we ensure that multiple threads running inside the enclave cannot alter each other's stack during an OCALL by removing that thread's stack from the isolated mapping.

\begin{tcolorbox}[boxsep=1pt,left=2pt,right=2pt,top=1pt,bottom=1pt,title={Summary}]
  We presented a microbenchmark evaluation of our SGX protection domain, showing the performance overhead for ending the isolation via a page fault or \syscall.
  The benchmarks demonstrate that \Approach outperforms previous solutions in this field without requiring additional hardware features.
  Our security evaluation showed that we improve the security of the system in the presence of a malicious enclave.
\end{tcolorbox}

\section{Discussion}\label{sec:discussion}
In this section, we discuss future work and related work.

\subsection{Future Work}\label{sec:future-work}
\paragraph{Enhanced \Syscall Filtering.}
The largest bottleneck for efficient deep argument filtering is the necessity of TLB flushes.
Future work can investigate the possibility of employing protection keys in the kernel, completely removing the necessity of modifying and flushing page tables.
Previous work has already explored the possibility of using such keys in the kernel~\cite{Gravani2019pkk}.
A more recent patch set has even provided protection-key functionality to the kernel~\cite{Corbet2020kernelpkk}.
As the reliance on protection keys contradicts the aim of this work for providing a method for commodity, off-the-shelve systems, they were not used.
Similarly, IMIX~\cite{Frassetto2018} could improve the performance of our deep argument filtering.
However, currently, no hardware supports IMIX.
Future work can investigate whether seccomp can be extended to include our approach directly.
To generate tight filters for complex arguments, the automated generation of \syscall filter rules~\cite{Ghavamnia2020,Canella2021chestnut,DeMarinis2020} can be extended to extract arguments automatically. 

\paragraph{SGX-Protection Domain.}
The benchmarks in \Cref{sec:sgx-eval:performance} show that the overhead between raising a page fault and handling it is quite high.
Merging the proof-of-concept directly into the kernel would increase the performance by greatly reducing the amount of code executed.

\subsection{Related Work}
\paragraph{\Syscall Filtering.}
Multiple works have been published on \syscall checking~\cite{Goldberg1996secure,Wagner1999,Provos2003,jain2000,garfinkel2004ostia,Acharya2000,Alexandrov1999,Peterson2002,Fraser2000,Lin2005,Dan1997,Ghormley1998}.
Some rely on kernel tracing~\cite{Goldberg1996secure,Wagner1999,Provos2003,jain2000,garfinkel2004ostia,Acharya2000,Alexandrov1999}, but the performance suffers from additional context switches.
Hence, Linux relies on seccomp to perform \syscall-filtering if it is requested by a developer.
As discussed, seccomp requires a developer to manually identify the \syscalls required by an application.
Several recent works have investigated the feasibility of automatically identifying and generating these filter rules, eliminating the need for manual analysis~\cite{DeMarinis2020,Ghavamnia2020,Canella2021chestnut}.
While seccomp improves the security, it negatively effects the performance (\cf \cref{sec:sandboxing-eval}).
Recent work has therefore proposed changing how seccomp handles filters to improve the performance~\cite{Hromatka2018,skarlatos2020}.
Nevertheless, deep argument filtering is still not supported as of Linux 5.11.
Sala{\"u}n~\cite{Salaun2020landlock} restricts ambient rights, such as global filesystem accesses, for a set of processes.
This is an orthogonal approach as it does not attempt to filter syscalls, but instead focuses on access control on kernel objects directly~\cite{Linux2020landlock}.

\paragraph{Memory Isolation.}
Memory isolation is a well-researched field where proposals can be grouped into OS-, \mbox{virtualization-,} hardware-, language- and runtime-based techniques.
While memory isolation can be easily achieved on the OS level by simply placing the necessary parts into separate processes, this does incur a significant overhead.
To prevent this significant overhead, recent work provides additional OS abstractions~\cite{Dautenhahn2015,Hsu2016,Litton2016} that together with compiler support~\cite{Chen2016shreds} or runtime analysis tools~\cite{Bittau2008} make it feasible to isolate long-term signing keys in a web server.
Other work has proposed to use a hypervisor for isolating memory.
Dune~\cite{Belay2012} allows a user-space process to use the Intel VT-x virtualization extensions to isolate compartments.
Other work demonstrated how the \textit{VMFUNC} instruction can be used to switch extended page tables to achieve in-process isolation~\cite{Koning2017,Liu2015Isolation}.
SIM~\cite{Sharif2009} uses VT-x for isolating a security monitor in an untrusted VM.

Memory isolation can also be enforced by the hardware, for instance, by Intel SGX or ARM TrustZone.
Previous work used SGX to protect internal data structures of just-in-time compilers to prevent code-injection attacks~\cite{Frassetto2017}.
While the hardware enforces this isolation, the switching overhead is similar to other approaches~\cite{Koning2017}.
Other works proposed additional x86 ISA extensions to add load and store instructions that must be used to access data in a safe region~\cite{Frassetto2018,Mogosanu2018}.
These approaches additionally require CFI~\cite{Abadi2005CFI} to protect against control-flow hijack attacks~\cite{Szekeres2013sok}.
Several works have relied on Intel Memory Protection Keys (MPK) to facilitate more efficient memory isolation~\cite{Vahldiek2018erim,Schrammel2020Donky,Koning2017,Hedayati2019hodor,Park2019libmpk}.
However, Intel MPK is not used in the kernel and not available for most commodity systems.
It is only available in a limited subset of shipped processors since the Intel Xeon Skylake microarchitecture.
Creative use of page-table entries for in-process isolation is a technique that has been explored~\cite{Frassetto2018,Koning2017}, but these solutions require modifications of the ISA or the re-purposing of ignored or reserved bits in page tables.
Consequently, they can only be applied to future hardware but not to commodity systems.

Memory isolation can also be ensured through static checks in memory-safe languages.
In unsafe languages, this isolation can be provided through software fault isolation, where runtime checks are added by the compiler or through binary rewriting~\cite{wahbe1993efficient,McCamant2006}.
Naturally, this imposes a performance overhead while not protecting against control-flow hijack attacks, making it necessary to combine it with CFI.
Recent work has explored the possibility of so-called \textit{zero-cost transitions} between normal and sandboxed code for well-structured code, but still requires CFI~\cite{Kolosick2021isolation}.
Contrary to previous solutions for memory isolation~\cite{Frassetto2018,Mogosanu2018,Vahldiek2018erim,Schrammel2020Donky,Koning2017,Hedayati2019hodor,Park2019libmpk}, \Approach does not require ISA extensions or re-purposing of ignored bits in the page table.
Instead, we solely rely on existing functionality and bring our enhanced filtering to widely available commodity systems.
\Approach retrofits sandboxing mechanisms with strategies that previous works have explored against microarchitectural attacks~\cite{Gruss2017KASLR,Hua2018epti}.

\paragraph{Intel SGX.}
The memory isolation of SGX is asymmetric, \ie while enclave memory is inaccessible for the operating system and the host application, the enclave has full access to the data and code of the host application.
Schwarz~\etal\cite{Schwarz2019SGXMalware} showed that untrusted SGX enclaves can rewrite the host application memory to impersonate the host and execute arbitrary \syscalls.
To mitigate such attacks, Weiser~\etal\cite{Weiser2019SGXJail} proposed SGXJail, which contains two ways to isolate enclaves, either isolation through another process or through a slightly modified variant of Intel MPK.
Other works have proposed to monitor the I/O behavior of enclaves to detect a potential attack~\cite{Davenport2014,Costan2016sgx}.
Another proposed approach is to analyze enclave code before running it, but this is not possible for generic loaders~\cite{Costan2016sgx}.
Hence, Costan and Devadas~\cite{Costan2016sgx} proposed to force generic loader enclaves to embed malware-analysis code into the enclave, but it is unknown how effective this approach is.

Other approaches try to improve the security of the enclave itself~\cite{Kuvaiskii2017,seo2017sgx,Zhao2020mptee}, but such approaches are orthogonal to our approach as they assume a malicious host or OS.
A SFI-based approach has been proposed by Ryoan~\cite{Hunt2016}, but this requires recompilation of the enclave using SFI, which might not be possible and is not necessary in our approach.

\section{Conclusion}\label{sec:conclusion}
In this paper, we proposed \Approach, a security-domain isolation mechanism for commodity off-the-shelf CPUs.
We presented two novel techniques for dynamic, time-limited changes to the memory isolation at security-critical points, called \ApproachFreeze and \ApproachHide.
While \ApproachFreeze makes memory temporarily read-only, \ApproachHide temporarily removes access permissions.
We evaluated the versatility of \Approach in two scenarios.
In the first scenario, \Approach enables faster and more fine-grained \syscall filtering than seccomp-bpf while providing stronger memory-safety guarantees that allow supporting deep argument filtering.
In the second scenario, \Approach efficiently confines SGX enclaves, outperforming existing solutions by \SI{14.6}{\percent}-\SI{22}{\percent} and removing hardware requirements. 
Our results show that \Approach is a viable mechanism that can be used to isolate domains within applications without relying on special hardware instructions or extensions. 

\section*{Acknowledgments}
This project has received funding from the European Research Council (ERC) under the European Union's Horizon 2020 research and innovation program (grant agreement No 681402).
Additional funding was provided by generous gifts from ARM, Amazon, and Intel.
Any opinions, findings, and conclusions or recommendations expressed in this paper are those of the authors and do not necessarily reflect the views of the funding parties.

\bibliographystyle{IEEEtranS}
\bibliography{main}

\begin{thebibliography}{10}
\providecommand{\url}[1]{#1}
\csname url@samestyle\endcsname
\providecommand{\newblock}{\relax}
\providecommand{\bibinfo}[2]{#2}
\providecommand{\BIBentrySTDinterwordspacing}{\spaceskip=0pt\relax}
\providecommand{\BIBentryALTinterwordstretchfactor}{4}
\providecommand{\BIBentryALTinterwordspacing}{\spaceskip=\fontdimen2\font plus
\BIBentryALTinterwordstretchfactor\fontdimen3\font minus
  \fontdimen4\font\relax}
\providecommand{\BIBforeignlanguage}[2]{{%
\expandafter\ifx\csname l@#1\endcsname\relax
\typeout{** WARNING: IEEEtranS.bst: No hyphenation pattern has been}%
\typeout{** loaded for the language `#1'. Using the pattern for}%
\typeout{** the default language instead.}%
\else
\language=\csname l@#1\endcsname
\fi
#2}}
\providecommand{\BIBdecl}{\relax}
\BIBdecl

\bibitem{Abadi2005CFI}
M.~Abadi, M.~Budiu, U.~Erlingsson, and J.~Ligatti, ``{Control-Flow
  Integrity},'' in \emph{{CCS}}, 2005.

\bibitem{Acharya2000}
A.~Acharya and M.~Raje, ``{MAPbox: Using Parameterized Behavior Classes to
  Confine Untrusted Applications},'' in \emph{{USENIX Security Symposium}},
  2000.

\bibitem{Alexandrov1999}
A.~Alexandrov, P.~Kmiec, and K.~Schauser, ``{Consh: Confined Execution
  Environment for Internet Computations},'' The University of California, Santa
  Barbara, Tech. Rep., 1999.

\bibitem{AndroidAppSandbox}
\BIBentryALTinterwordspacing
Android, ``{Application Sandbox},'' 2021. [Online]. Available:
  \url{https://source.android.com/security/app-sandbox}
\BIBentrySTDinterwordspacing

\bibitem{AppArmor}
\BIBentryALTinterwordspacing
AppArmor, ``{AppArmor: Linux kernel security module},'' 2021. [Online].
  Available: \url{https://apparmor.net/}
\BIBentrySTDinterwordspacing

\bibitem{Backes2014oxymoron}
M.~Backes and S.~N{\"u}rnberger, ``Oxymoron: Making fine-grained memory
  randomization practical by allowing code sharing,'' in \emph{{USENIX Security
  Symposium}}, 2014.

\bibitem{Belay2012}
A.~Belay, A.~Bittau, A.~Mashtizadeh, D.~Terei, D.~Mazi{\`e}res, and
  C.~Kozyrakis, ``{Dune: Safe User-level Access to Privileged CPU Features},''
  in \emph{{OSDI}}, 2012.

\bibitem{Bittau2008}
A.~Bittau, P.~Marchenko, M.~Handley, and B.~Karp, ``{Wedge: Splitting
  Applications into Reduced-Privilege Compartments},'' in \emph{{NSDI}}, 2008.

\bibitem{Bueso2019getppid}
\BIBentryALTinterwordspacing
D.~Bueso, ``{tools/perf-bench: Add basic syscall benchmark},'' 2019. [Online].
  Available: \url{https://lore.kernel.org/patchwork/patch/1048777/}
\BIBentrySTDinterwordspacing

\bibitem{Canella2021chestnut}
C.~Canella, M.~Werner, D.~Gruss, and M.~Schwarz, ``{Automating Seccomp Filter
  Generation for Linux Applications},'' in \emph{{CCSW}}, 2021.

\bibitem{Checkoway2010JOP}
S.~Checkoway, L.~Davi, A.~Dmitrienko, A.~Sadeghi, H.~Shacham, and M.~Winandy,
  ``Return-oriented programming without returns,'' in \emph{{CCS}}, 2010.

\bibitem{Chen2016shreds}
Y.~Chen, S.~Reymondjohnson, Z.~Sun, and L.~Lu, ``{Shreds: Fine-Grained
  Execution Units with Private Memory},'' in \emph{{S\&P}}, 2016.

\bibitem{Corbet2020kernelpkk}
\BIBentryALTinterwordspacing
J.~Corbet, ``{Memory protection keys for the kernel},'' 2020. [Online].
  Available: \url{https://lwn.net/Articles/826554/}
\BIBentrySTDinterwordspacing

\bibitem{Costan2016sgx}
V.~Costan and S.~Devadas, ``{Intel SGX Explained},'' \emph{Cryptology ePrint
  Archive, Report 2016/086}, 2016.

\bibitem{Dan1997}
A.~Dan, A.~Mohindra, R.~Ramaswami, and D.~Sitaram, ``{Chakravyuha (CV): A
  sandbox operating system environment for controlled execution of alien
  code},'' Tech. Rep., 1997.

\bibitem{Dautenhahn2015}
N.~Dautenhahn, T.~Kasampalis, W.~Dietz, J.~Criswell, and V.~Adve, ``{Nested
  Kernel: An Operating System Architecture for Intra-Kernel Privilege
  Separation},'' in \emph{{ASPLOS}}, 2015.

\bibitem{Davenport2014}
\BIBentryALTinterwordspacing
S.~Davenport and R.~Ford, ``{SGX: the good, the bad and the downright ugly},''
  January 2014. [Online]. Available:
  \url{https://www.virusbulletin.com/virusbulletin/2014/01/sgx-good-bad-and-downright-ugly}
\BIBentrySTDinterwordspacing

\bibitem{DeMarinis2020}
N.~DeMarinis, K.~Williams-King, D.~Jin, R.~Fonseca, and V.~P. Kemerlis,
  ``{sysfilter: Automated System Call Filtering for Commodity Software},'' in
  \emph{RAID}, 2020.

\bibitem{Linux2020landlock}
\BIBentryALTinterwordspacing
K.~development community, ``{Landlock: kernel documentation},'' 2020. [Online].
  Available:
  \url{https://landlock.io/linux-doc/landlock-v21/security/landlock/kernel.html}
\BIBentrySTDinterwordspacing

\bibitem{Edge2015seccomp}
\BIBentryALTinterwordspacing
J.~Edge, ``{A seccomp overview},'' 2015. [Online]. Available:
  \url{https://lwn.net/Articles/656307/}
\BIBentrySTDinterwordspacing

\bibitem{Edge2019}
\BIBentryALTinterwordspacing
------, ``{Deep argument inspection for seccomp},'' September 2019. [Online].
  Available: \url{https://lwn.net/Articles/799557/}
\BIBentrySTDinterwordspacing

\bibitem{Edge2020}
\BIBentryALTinterwordspacing
------, ``{Seccomp and deep argument inspection},'' June 2020. [Online].
  Available: \url{https://lwn.net/Articles/822256/}
\BIBentrySTDinterwordspacing

\bibitem{Ford2008}
B.~Ford and R.~Cox, ``{Vx32: Lightweight User-level Sandboxing on the x86},''
  in \emph{Usenix ATC}, 2008.

\bibitem{Fraser2000}
T.~Fraser, L.~Badger, and M.~Feldman, ``{Hardening COTS software with generic
  software wrappers},'' in \emph{{DISCEX}}, 2000.

\bibitem{Frassetto2017}
T.~Frassetto, D.~Gens, C.~Liebchen, and A.-R. Sadeghi, ``{JITGuard: Hardening
  Just-in-Time Compilers with SGX},'' in \emph{{CCS}}, 2017.

\bibitem{Frassetto2018}
T.~Frassetto, P.~Jauernig, C.~Liebchen, and A.-R. Sadeghi, ``{IMIX}: In-process
  memory isolation extension,'' in \emph{{USENIX Security Symposium}}, 2018.

\bibitem{garfinkel2003traps}
T.~Garfinkel, ``{Traps and Pitfalls: Practical Problems in System Call
  Interposition Based Security Tools},'' in \emph{{NDSS}}, 2003.

\bibitem{garfinkel2004ostia}
T.~Garfinkel, B.~Pfaff, M.~Rosenblum \emph{et~al.}, ``Ostia: A delegating
  architecture for secure system call interposition.'' in \emph{NDSS}, 2004.

\bibitem{Ghavamnia2020}
S.~Ghavamnia, T.~Palit, S.~Mishra, and M.~Polychronakis, ``{Temporal System
  Call Specialization for Attack Surface Reduction},'' in \emph{{USENIX
  Security Symposium}}, 2020.

\bibitem{Ghormley1998}
D.~P. Ghormley, D.~Petrou, S.~H. Rodrigues, and T.~E. Anderson, ``{SLIC: An
  Extensibility System for Commodity Operating Systems},'' in \emph{{USENIX
  ATC}}, 1998.

\bibitem{Goktas2014COP}
E.~G{\"{o}}ktas, E.~Athanasopoulos, H.~Bos, and G.~Portokalidis, ``Out of
  control: Overcoming control-flow integrity,'' in \emph{{S\&P}}, 2014.

\bibitem{Goldberg1996secure}
I.~Goldberg, D.~Wagner, R.~Thomas, E.~A. Brewer \emph{et~al.}, ``A secure
  environment for untrusted helper applications: Confining the wily hacker,''
  in \emph{USENIX Security Symposium}, 1996.

\bibitem{Gravani2019pkk}
S.~Gravani, M.~Hedayati, J.~Criswell, and M.~L. Scott, ``{IskiOS: Lightweight
  Defense Against Kernel-Level Code-Reuse Attacks},'' \emph{arXiv:1903.04654},
  2019.

\bibitem{Gruss2017KASLR}
D.~Gruss, M.~Lipp, M.~Schwarz, R.~Fellner, C.~Maurice, and S.~Mangard, ``{KASLR
  is Dead: Long Live KASLR},'' in \emph{{ESSoS}}, 2017.

\bibitem{Hedayati2019hodor}
M.~Hedayati, S.~Gravani, E.~Johnson, J.~Criswell, M.~L. Scott, K.~Shen, and
  M.~Marty, ``{Hodor: Intra-Process Isolation for High-Throughput Data Plane
  Libraries},'' in \emph{{Usenix ATC}}, 2019.

\bibitem{Hromatka2018}
T.~Hromatka, ``{seccomp and libseccomp performance improvements},'' 2018.

\bibitem{Hsu2016}
T.~C.-H. Hsu, K.~Hoffman, P.~Eugster, and M.~Payer, ``{Enforcing Least
  Privilege Memory Views for Multithreaded Applications},'' in \emph{{CCS}},
  2016.

\bibitem{Hua2018epti}
Z.~Hua, D.~Du, Y.~Xia, H.~Chen, and B.~Zang, ``{EPTI: efficient defence against
  meltdown attack for unpatched VMs},'' in \emph{{USENIX ATC}}, 2018.

\bibitem{Hunt2016}
T.~Hunt, Z.~Zhu, Y.~Xu, S.~Peter, and E.~Witchel, ``Ryoan: {A} distributed
  sandbox for untrusted computation on secret data,'' in \emph{Usenix OSDI},
  2016.

\bibitem{Android2017seccomp}
\BIBentryALTinterwordspacing
G.~Inc., ``{Seccomp filter in Android O},'' 2017. [Online]. Available:
  \url{https://android-developers.googleblog.com/2017/07/seccomp-filter-in-android-o.html}
\BIBentrySTDinterwordspacing

\bibitem{Intel_vol3}
{Intel}, ``{Intel 64 and IA-32 Architectures Software Developer's Manual,
  Volume 3 (3A, 3B \& 3C): System Programming Guide},'' 2019.

\bibitem{Ispoglou2018}
K.~K. Ispoglou, B.~AlBassam, T.~Jaeger, and M.~Payer, ``{Block Oriented
  Programming: Automating Data-Only Attacks},'' in \emph{CCS}, 2018.

\bibitem{jain2000}
K.~Jain and R.~Sekar, ``{User-Level Infrastructure for System Call
  Interposition: A Platform for Intrusion Detection and Confinement}.''\hskip
  1em plus 0.5em minus 0.4em\relax {NDSS}, 2000.

\bibitem{Kemerlis2015}
V.~Kemerlis, ``Protecting commodity operating systems through strong kernel
  isolation,'' Ph.D. dissertation, Columbia University, 2015.

\bibitem{Kemerlis2014}
V.~P. Kemerlis, M.~Polychronakis, and A.~D. Keromytis, ``ret2dir: Rethinking
  kernel isolation,'' in \emph{USENIX Security Symposium}, 2014.

\bibitem{Kemerlis2012}
V.~P. Kemerlis, G.~Portokalidis, and A.~D. Keromytis, ``kguard: Lightweight
  kernel protection against return-to-user attacks,'' in \emph{{USENIX}
  Security Symposium}, 2012.

\bibitem{Kenjar2020v0ltpwn}
Z.~Kenjar, T.~Frassetto, D.~Gens, M.~Franz, and A.~Sadeghi, ``{V0LTpwn:
  Attacking x86 Processor Integrity from Software},'' in \emph{USENIX Security
  Symposium}, 2020.

\bibitem{Kim2014}
Y.~Kim, R.~Daly, J.~Kim, C.~Fallin, J.~H. Lee, D.~Lee, C.~Wilkerson, K.~Lai,
  and O.~Mutlu, ``{Flipping Bits in Memory Without Accessing Them: An
  Experimental Study of DRAM Disturbance Errors},'' in \emph{ISCA}, 2014.

\bibitem{Kiriansky2018speculative}
V.~Kiriansky and C.~Waldspurger, ``{Speculative Buffer Overflows: Attacks and
  Defenses},'' \emph{arXiv:1807.03757}, 2018.

\bibitem{Kolosick2021isolation}
M.~Kolosick, S.~Narayan, C.~Watt, M.~LeMay, D.~Garg, R.~Jhala, and D.~Stefan,
  ``{Isolation Without Taxation: Near Zero Cost Transitions for SFI},''
  \emph{arXiv:2105.00033}, 2021.

\bibitem{Koning2017}
K.~Koning, X.~Chen, H.~Bos, C.~Giuffrida, and E.~Athanasopoulos, ``{No Need to
  Hide: Protecting Safe Regions on Commodity Hardware},'' in \emph{{EuroSys}},
  2017.

\bibitem{Kuvaiskii2017}
D.~Kuvaiskii, O.~Oleksenko, S.~Arnautov, B.~Trach, P.~Bhatotia, P.~Felber, and
  C.~Fetzer, ``{SGXBOUNDS: Memory Safety for Shielded Execution},'' in
  \emph{{EuroSys}}, 2017.

\bibitem{Kuznetsov2014CPI}
V.~Kuznetsov, L.~Szekeres, M.~Payer, G.~Candea, R.~Sekar, and D.~Song,
  ``{Code-Pointer Integrity},'' in \emph{{OSDI}}, 2014.

\bibitem{Lan2015loop}
B.~Lan, Y.~Li, H.~Sun, C.~Su, Y.~Liu, and Q.~Zeng, ``Loop-oriented programming:
  a new code reuse attack to bypass modern defenses,'' in \emph{IEEE
  Trustcom/BigDataSE/ISPA}, 2015.

\bibitem{Larabel2021tlb}
\BIBentryALTinterwordspacing
M.~Larabel, ``{Concurrent TLB Flushing For Linux 5.13 Provide A Small
  Performance Benefit},'' 2021. [Online]. Available:
  \url{https://www.phoronix.com/scan.php?page=news_item&px=Linux-5.13-Concurrent-TLB-Flush}
\BIBentrySTDinterwordspacing

\bibitem{Lin2005}
C.~M. Linn, M.~Rajagopalan, S.~Baker, C.~Collberg, S.~K. Debray, and J.~H.
  Hartman, ``{Protecting against Unexpected System Calls},'' in \emph{{USENIX
  Security Symposium}}, 2005.

\bibitem{Lipp2018meltdown}
M.~Lipp, M.~Schwarz, D.~Gruss, T.~Prescher, W.~Haas, A.~Fogh, J.~Horn,
  S.~Mangard, P.~Kocher, D.~Genkin, Y.~Yarom, and M.~Hamburg, ``{Meltdown:
  Reading Kernel Memory from User Space},'' in \emph{USENIX Security
  Symposium}, 2018.

\bibitem{Litton2016}
J.~Litton, A.~Vahldiek-Oberwagner, E.~Elnikety, D.~Garg, B.~Bhattacharjee, and
  P.~Druschel, ``Light-weight contexts: An {OS} abstraction for safety and
  performance,'' in \emph{{OSDI 16}}, 2016.

\bibitem{Liu2015Isolation}
Y.~Liu, T.~Zhou, K.~Chen, H.~Chen, and Y.~Xia, ``{Thwarting Memory Disclosure
  with Efficient Hypervisor-Enforced Intra-Domain Isolation},'' in
  \emph{{CCS}}, 2015.

\bibitem{Lu2015aslrguard}
K.~Lu, C.~Song, B.~Lee, S.~P. Chung, T.~Kim, and W.~Lee, ``Aslr-guard: Stopping
  address space leakage for code reuse attacks,'' in \emph{CCS}, 2015.

\bibitem{LWN_kpti}
\BIBentryALTinterwordspacing
LWN, ``The current state of kernel page-table isolation,'' 2017. [Online].
  Available: \url{https://lwn.net/SubscriberLink/741878/eb6c9d3913d7cb2b/}
\BIBentrySTDinterwordspacing

\bibitem{Matousek2016procmem}
\BIBentryALTinterwordspacing
P.~Matousek, ``{CVE-2016-5195 kernel: mm: privilege escalation via MAP\_PRIVATE
  COW breakage},'' 2016. [Online]. Available:
  \url{https://bugzilla.redhat.com/show_bug.cgi?id=1384344#c16}
\BIBentrySTDinterwordspacing

\bibitem{Matz2014abi}
M.~Matz, J.~Hubicka, A.~Jaeger, and M.~Mitchell, ``{System V Application Binary
  Interface},'' \emph{AMD64 Architecture Processor Supplement, Draft v0.99.7},
  2014.

\bibitem{McCamant2006}
S.~McCamant and G.~Morrisett, ``{Evaluating SFI for a CISC Architecture},'' in
  \emph{{USENIX Security Symposium}}, 2006.

\bibitem{Mogosanu2018}
L.~Mogosanu, A.~Rane, and N.~Dautenhahn, ``{MicroStache: A Lightweight
  Execution Context for In-Process Safe Region Isolation},'' in \emph{RAID},
  2018.

\bibitem{Murdock2019plundervolt}
K.~Murdock, D.~Oswald, F.~D. Garcia, J.~Van~Bulck, D.~Gruss, and F.~Piessens,
  ``{Plundervolt: Software-based Fault Injection Attacks against Intel SGX},''
  in \emph{{S\&P}}, 2020.

\bibitem{Nergal2001ret2libc}
Nergal, ``{The advanced return-into-lib(c) explits: PaX case study},'' 2001.

\bibitem{Park2019libmpk}
S.~Park, S.~Lee, W.~Xu, H.~Moon, and T.~Kim, ``{libmpk: Software Abstraction
  for Intel Memory Protection Keys (Intel {MPK})},'' in \emph{{USENIX ATC}},
  2019.

\bibitem{Peterson2002}
D.~S. Peterson, M.~Bishop, and R.~Pandey, ``{A Flexible Containment Mechanism
  for Executing Untrusted Code},'' in \emph{{USENIX Security Symposium}}, 2002.

\bibitem{Prevelakis2001sandboxing}
V.~Prevelakis and D.~Spinellis, ``Sandboxing applications,'' in \emph{USENIX
  ATC}, 2001.

\bibitem{Provos2003}
N.~Provos, ``Improving host security with system call policies,'' in
  \emph{{USENIX Security Symposium}}, 2003.

\bibitem{sgxbench}
\BIBentryALTinterwordspacing
R.~Quinonez, ``{SGXBENCH framework for benchmarking SGX enclaves},'' 2018.
  [Online]. Available: \url{https://github.com/sgxbench/sgxbench}
\BIBentrySTDinterwordspacing

\bibitem{Reis2019SiteIsolation}
C.~Reis, A.~Moshchuk, and N.~Oskov, ``{Site Isolation: Process Separation for
  Web Sites within the Browser},'' in \emph{USENIX Security Symposium}, 2019.

\bibitem{Rogowski2017}
R.~Rogowski, M.~Morton, F.~Li, F.~Monrose, K.~Z. Snow, and M.~Polychronakis,
  ``{Revisiting Browser Security in the Modern Era: New Data-Only Attacks and
  Defenses},'' in \emph{{EuroS\&P}}, 2017.

\bibitem{Salaun2020landlock}
M.~Sala{\"u}n, ``{Landlock: unprivileged access control},'' 2016.

\bibitem{Schrammel2020Donky}
D.~Schrammel, S.~Weiser, S.~Steinegger, M.~Schwarzl, M.~Schwarz, S.~Mangard,
  and D.~Gruss, ``{Donky: Domain Keys--Efficient In-Process Isolation for
  RISC-V and x86},'' in \emph{USENIX Security Symposium}, 2020.

\bibitem{Schuster2015COOP}
F.~Schuster, T.~Tendyck, C.~Liebchen, L.~Davi, A.~Sadeghi, and T.~Holz,
  ``{Counterfeit Object-oriented Programming: On the Difficulty of Preventing
  Code Reuse Attacks in {C++} Applications},'' in \emph{{S\&P}}, 2015.

\bibitem{Schwarz2019STL}
M.~Schwarz, C.~Canella, L.~Giner, and D.~Gruss, ``{Store-to-Leak Forwarding:
  Leaking Data on Meltdown-resistant CPUs},'' \emph{arXiv:1905.05725}, 2019.

\bibitem{Schwarz2018DF}
M.~Schwarz, D.~Gruss, M.~Lipp, C.~Maurice, T.~Schuster, A.~Fogh, and
  S.~Mangard, ``{Automated Detection, Exploitation, and Elimination of
  Double-Fetch Bugs using Modern CPU Features},'' in \emph{AsiaCCS}, 2018.

\bibitem{SchwarzPteditor}
\BIBentryALTinterwordspacing
M.~Schwarz, M.~Lipp, and C.~Canella, ``{misc0110/PTEditor: A small library to
  modify all page-table levels of all processes from user space for x86\_64 and
  ARMv8},'' 2018. [Online]. Available:
  \url{https://github.com/misc0110/PTEditor}
\BIBentrySTDinterwordspacing

\bibitem{Schwarz2019SGXMalware}
M.~Schwarz, S.~Weiser, and D.~Gruss, ``{Practical Enclave Malware with Intel
  SGX},'' in \emph{DIMVA}, 2019.

\bibitem{seo2017sgx}
J.~Seo, B.~Lee, S.~M. Kim, M.-W. Shih, I.~Shin, D.~Han, and T.~Kim,
  ``Sgx-shield: Enabling address space layout randomization for sgx programs.''
  in \emph{NDSS}, 2017.

\bibitem{Shacham2007}
H.~Shacham, ``{The geometry of innocent flesh on the bone: Return-into-libc
  without function calls (on the x86)},'' in \emph{CCS}, 2007.

\bibitem{Sharif2009}
M.~I. Sharif, W.~Lee, W.~Cui, and A.~Lanzi, ``{Secure In-VM Monitoring Using
  Hardware Virtualization},'' in \emph{{CCS}}, 2009.

\bibitem{skarlatos2020}
D.~Skarlatos, Q.~Chen, J.~Chen, T.~Xu, and J.~Torrellas, ``{Draco:
  Architectural and Operating System Support for System Call Security},'' in
  \emph{{MICRO}}, 2020.

\bibitem{Szekeres2013sok}
L.~Szekeres, M.~Payer, T.~Wei, and D.~Song, ``{SoK: Eternal War in Memory},''
  in \emph{S\&P}, 2013.

\bibitem{Tizen2014}
\BIBentryALTinterwordspacing
Tizen, ``{Security:Seccomp},'' 2018. [Online]. Available:
  \url{https://wiki.tizen.org/Security:Seccomp}
\BIBentrySTDinterwordspacing

\bibitem{Vahldiek2018erim}
A.~Vahldiek-Oberwagner, E.~Elnikety, N.~O. Duarte, M.~Sammler, P.~Druschel, and
  D.~Garg, ``{ERIM: Secure and Efficient In-process Isolation with Memory
  Protection Keys},'' in \emph{{USENIX Security Symposium}}, 2019.

\bibitem{Vanbulck2020lvi}
J.~Van~Bulck, D.~Moghimi, M.~Schwarz, M.~Lipp, M.~Minkin, D.~Genkin, Y.~Yuval,
  B.~Sunar, D.~Gruss, and F.~Piessens, ``{LVI: Hijacking Transient Execution
  through Microarchitectural Load Value Injection},'' in \emph{S\&P}, 2020.

\bibitem{Wagner1999}
D.~A. Wagner, ``{Janus: An Approach for Confinement of Untrusted
  Applications},'' University of California at Berkeley, Tech. Rep., 1999.

\bibitem{wahbe1993efficient}
R.~Wahbe, S.~Lucco, T.~E. Anderson, and S.~L. Graham, ``{Efficient
  software-based fault isolation},'' in \emph{ACM SOSP}, 1993.

\bibitem{Wang2017DF}
P.~Wang and J.~Krinke, ``{How Double-Fetch Situations turn into Double-Fetch
  Vulnerabilities: A Study of Double Fetches in the Linux Kernel},'' in
  \emph{{USENIX Security Symposium}}, 2017.

\bibitem{Watson2007}
R.~N.~M. Watson, ``{Exploiting Concurrency Vulnerabilities in System Call
  Wrappers},'' in \emph{WOOT}, 2007.

\bibitem{Weiser2019SGXJail}
S.~Weiser, L.~Mayr, M.~Schwarz, and D.~Gruss, ``{SGXJail}: Defeating enclave
  malware via confinement,'' in \emph{RAID}, 2019.

\bibitem{Firefox2019fission}
\BIBentryALTinterwordspacing
M.~Wiki, ``{Project Fission},'' 2019. [Online]. Available:
  \url{https://wiki.mozilla.org/Project_Fission}
\BIBentrySTDinterwordspacing

\bibitem{Firefox2019sandbox}
\BIBentryALTinterwordspacing
------, ``{Security/Sandbox},'' 2019. [Online]. Available:
  \url{https://wiki.mozilla.org/Security/Sandbox}
\BIBentrySTDinterwordspacing

\bibitem{SELinuxFAQ}
\BIBentryALTinterwordspacing
S.~Wiki, ``{FAQ --- SELinux Wiki},'' 2009. [Online]. Available:
  \url{http://selinuxproject.org/w/?title=FAQ&oldid=729}
\BIBentrySTDinterwordspacing

\bibitem{Zhao2020mptee}
W.~Zhao, K.~Lu, Y.~Qi, and S.~Qi, ``{MPTEE: Bringing Flexible and Efficient
  Memory Protection to Intel SGX},'' in \emph{EuroSys}, 2020.

\end{thebibliography}

\appendix

\begin{table*}[t]
  \centering
  \caption{The specific commands which were used for evaluating the performance of an unsandboxed, seccomp-, and \Approach-sandboxed version of the respective application.}
  \label{tbl:real-world-commands}
  \adjustbox{max width=\hsize}{
    \begin{tabular}{llc}
       & \textbf{Software} & \textbf{Command}                                                                     \\
      \toprule
      \multirow{9}{*}{busybox}
       & diff              & busybox diff cat.sh grep.sh                                                          \\
       & true              & busybox true                                                                         \\
       & env               & busybox env                                                                          \\
       & ls                & busybox ls                                                                           \\
       & dmesg             & busybox dmesg                                                                        \\
       & cat               & busybox cat test                                                                     \\
       & head              & busybox head -n 100 test                                                             \\
       & grep              & busybox grep -ir python3 .                                                           \\
       & pwd               & busybox pwd                                                                          \\
      \midrule
      \multirow{2}{*}{git}
       & diff              & git diff                                                                             \\
       & status            & git status                                                                           \\
      \midrule
      \multirow{6}{*}{ffmpeg}
       & extract           & ffmpeg\_g -i video.mp4 -r 1 -f image2 image-\%2d.png -y -hide\_banner                 \\
       & convert           & ffmpeg\_g -i video.mp4 video.avi -y -hide\_banner                                     \\
       & remove            & ffmpeg\_g -i video.mp4 -an output.mp4 -y -hide\_banner                                \\
       & crop              & ffmpeg\_g -i video.mp4 -filter:v 'crop=640:480:200:150' output.mp4 -y -hide\_banner   \\
       & info              & ffmpeg\_g -hide\_banner -i video.mp4                                                  \\
       & change            & ffmpeg\_g -i video.mp4 -filter:v scale=1280:720 -c:a copy output.mp4 -y -hide\_banner \\
      \bottomrule
    \end{tabular}
  }
\end{table*}

\section{Real-world Application Commands}\label{app:commands}
In \cref{tbl:real-world-commands}, we show the exact commands that were used for the evaluation of the real-world applications in \cref{sec:sandboxing-eval}.

\end{document}